\documentclass[twocolumn, iop, apj, twocolappendix, numberedappendix, appendixfloats]{emulateapj}
\usepackage{CJKutf8}
\usepackage{threeparttable}
\usepackage{multirow}
\usepackage{times}
\usepackage{enumitem}
\usepackage{url}
\usepackage{color}
\usepackage{amsmath,bm}
\usepackage{subfigure}
\usepackage{natbib}
\usepackage{graphicx}
\usepackage{graphics}
\usepackage{hyperref}
\usepackage{amsmath}
\usepackage{bm}
\usepackage{subfigure}
\usepackage{natbib}
\usepackage{array}
\usepackage{booktabs}
\usepackage{lineno}

\shorttitle{Mapping the Chemo-dynamics of the Galactic disk}
\shortauthors{Sun et al.}
\submitted{Submitted to ApJS;\,\,\,\,Accepted March 03, 2024}

\begin{document}
%\linenumbers

\title{Mapping the Chemo-dynamics of the Galactic disk using the LAMOST and APOGEE red clump stars}

\author{Weixiang Sun\textsuperscript{1,3}}
\author{Han Shen\textsuperscript{2}}
\author{Biwei Jiang\textsuperscript{1,3}}
\author{Xiaowei Liu\textsuperscript{2,3}}

\altaffiltext{1}{Department of Astronomy, Beijing Normal University, Beijing 100875, People's Republic of China; {\it sunweixiang@bnu.edu.cn (WXS)}; {\it bjiang@bnu.edu.cn (BWJ)}.}

\altaffiltext{2}{South-Western Institute for Astronomy Research, Yunnan University, Kunming 650500, People's Republic of China; {\it x.liu@ynu.edu.cn (XWL)}.}

%\altaffiltext{3}{Tianjin Astrophysics Center, Tianjin Normal University, Tianjin 300387, People’s Republic of China}

%\altaffiltext{4}{Department of Astronomy, Yunnan University, Kunming 650500, People's Republic of China;}

%\altaffiltext{5}{LAMOST Fellow}

\altaffiltext{3}{Corresponding authors}

\begin{abstract}
A detailed measurement is made of the metallicity distributions, kinematics and dynamics of the thin and thick disks, across a large disk volume (5.0 $\leq$ $R$ $\leq$ 15.0\,kpc and $|Z|$ $\leq$3.0\,kpc), by using the LAMOST-APOGEE red clump stars.
The metallicity distributions results show that the radial metallicity gradient $\Delta$[Fe/H]/$\Delta$R of the thin disk weakens with $|Z|$ from $-$0.06\,dex\,kpc$^{-1}$ at around $|Z|$ $<$ 0.25\,kpc to $-$0.02\,dex\,kpc$^{-1}$ at around $|Z|$ $>$ 2.75\,kpc, while the thick disk displays a global weak positive $\Delta$[Fe/H]/$\Delta$R, generally weaker than 0.01\,dex\,kpc$^{-1}$.
The vertical metallicity gradient $\Delta$[Fe/H]/$\Delta|Z|$ weakened steadily from $-$0.36\,dex\,kpc$^{-1}$ at $R$ $\sim$ 5.5\,kpc to $-$0.05\,dex\,kpc$^{-1}$ at around R $>$ 11.5\,kpc for the thin disk, while the thick disk presents an almost constant value (nearly $-$0.06 $\sim$ $-$0.08\,dex\,kpc$^{-1}$) for all the $R$ bins.
These results indicate the contribution of the radial migration to the disk evolution, and the obvious north-south asymmetry in [Fe/H] may be linked to the disk warp and/or the disk perturbation events.
The oscillations of the corrected $\Delta$[Fe/H]/$\Delta|Z|$ with $R$ are likely because of the resonances with the Galactic Bar.
Our detailed measurements of $\Delta$V$_{\phi}$/$\Delta$[Fe/H] indicate an ``inside-out" and ``upside-down" star formation scenario for the thick disk.
The results of eccentricity distributions and [$\alpha$/Fe]--velocity dispersion relations are likely to suggest that the thick disk stars require an obvious contribution from other heating mechanisms such as merger and accretion, or born in the chaotic mergers of gas-rich systems and/or turbulent interstellar medium.

\end{abstract}

\keywords{Stars: abundance -- Stars: kinematics and dynamics -- Galaxy: disk -- Galaxy: formation and evolution}

\section{Introduction}

Exploring the nature of the Galactic disk plays an important role in understanding the formation and evolution of the disk galaxies \citep[e.g.,][]{Yoshii1982, Hartkopf1982, Bensby2005, Pilkington2012, Kordopatis2015, Gaia Collaboration2023a, Gaia Collaboration2023b}.
Since the first discovered the thin and thick disks of the Milky Way by fitting the vertical stellar density profile with two exponential functions \citep[e.g,][]{Yoshii1982, Gilmore1983}, the idea of working on two independent disk components of the Galaxy has become the most mainstream and constructive idea in Galactic archaeology.

The thin and thick disks are not only different in spatial structure \citep[e.g.,][]{Du2003, Bensby2005, Chang2011}, but also in chemistry, age and kinematics, as well as chemo-dynamics \citep[e.g.,][]{Fuhrmann1998, Lee2011, Bland-Hawthorn2016, Miranda2016, Mackereth2017, Sun2020, Sun2023}.

Spatially, the scale-heights of thin and thick disks are about 0.2 $\sim$ 0.4\,kpc and 0.6 $\sim$ 1.0\,kpc \citep[e.g.,][]{Du2006, Juric2008, Bland-Hawthorn2016}, respectively.
The scale-lengths of thin and thick disks are about 1.0 $\sim$ 4.7\,kpc and 2.0 $\sim$ 5.5\,kpc \citep[e.g.,][]{Bilir2006, Karaali2007, Jia2014, Mackereth2017}, respectively.

Regards to chemistry and age, the thin disk stars generally have richer metallicity ([Fe/H]), lower $\alpha$-abundance ratio ([$\alpha$/Fe]), and younger ages than those of the thick disk stars \citep[e.g.,][]{Fuhrmann2008, Haywood2013, Vickers2021, Gent2022, Lian2022}, with the [Fe/H] distribution function of thin and thick disks respectively peaking around $-$0.2\,dex and $-$0.5 $\sim$ $-$0.6\,dex \citep[e.g,][]{Wyse1995, Soubiran2003, Kordopatis2011, Hayden2015, Sun2020}, and the [$\alpha$/Fe] is lower by 0.2 $\sim$ 0.3\,dex for the thin disk stars than that for the thick disk stars \citep[e.g.,][]{Lee2011, Adibekyan2013, Boeche2013, Anders2014, Yan2019}.
The thin disk generally has a negative radial metallicity gradient  \citep[e.g.,][]{Bilir2012, Pilkington2012, Donor2020, Katz2021, Imig2023}, whereas the thick disk has no or positive radial metallicity gradient \citep[e.g.,][]{Recio-Blanco2014, Coskunoglu2012, Miranda2016}.
For the vertical metallicity gradient, some results suggested that both thin and thick disks have negative values \citep[e.g.,][]{Chen2011, Carrell2012, Li2017, Tuncel2019}, while other results tend to suggest a flat gradient for the thick disk \citep[e.g.,][]{Katz2011, Li2018}.

Kinematically and dynamically, the thin disk stars generally have larger orbital rotation velocity, lower velocity dispersion, lower eccentricity and lower maximum distance from the Galactic plane ($Z_{max}$) than those of the thick disk stars \citep[e.g.,][]{Lee2011, Jing2016, Mackereth2019, Robin2022}.
The metallicity gradients of orbital rotation velocities are negative and positive for thin and thick disks \citep[e..g,][]{Lee2011, Han2020}, respectively.
The metallicity gradients of eccentricities are positive and negative for thin and thick disks \citep[e.g.,][]{Lee2011, Yan2019}, respectively.

All these distinguished properties for the two disks mean that the thin and thick disks may have experienced different formation mechanisms and evolution histories \citep[e.g.,][]{Jenkins1992, Brook2004, Schonrich2017, Mackereth2019, Han2020}.

Spitzer \& Schwarzschild ({\color{blue}{1953}}) suggested that old stars will become kinematically hotter with large velocity dispersion as they are scattered by the giant molecular clouds (GMCs).
Other studies also revealed that the spiral arms might contribute to their larger velocity dispersion \citep{Barbanis1967}.
However, Jenkins ({\color{blue}{1992}}) reported that when fully considering the heating of the GMCs and spiral arms, the acceptable heating constant is not enough to explain so kinematically hot properties of the observed old stars.
Therefore, the suggestions of violent origins of the thick disk are conceived, including:
(i) the thick disk stars formed from heating pre-existing thin disk stars by satellite mergers \citep[e.g.,][]{Quinn1993, Villalobos2008};
(ii) the thick disk stars formed by member stars that accreted from satellite galaxies during the merging process \citep[e.g.,][]{Abadi2003};
(iii) the thick disk stars formed in an environment with chaotic mergers of gas-rich system \citep[e.g.,][]{Brook2004, Brook2005, Brook2007, Bournaud2009}.
Recently, studies reported that the inside-out and upside-down star formation \citep[e.g.,][]{Kawata2017, Schonrich2017, Bird2021}, along with a cumulative secular process with stellar migration scenario by churning and blurring \citep[e.g.,][]{Loebman2011, Minchev2012, Roskar2012, Kordopatis2015, Schonrich2017} can also reshape a thick disk component of our Galaxy.

Previous studies \citep[e.g.,][]{Lee2011, Mackereth2019, Han2020} suggested that a detailed measurement of chemical, kinematic and dynamic properties of the thin/thick disk is useful for constraining the origin of these stars.
However, the observation limitation \citep[e.g.,][]{Lee2011, Bensby2014, Mackereth2019, Han2020} meant that these properties are not yet well measured in a larger Galactic disk radius which would make a credible assessment of the origins of the thick disk stars.
At present, the red clump (RC) stars \citep[e.g.,][]{Bovy2014, Huang2020} from the APOGEE \citep{Majewski2017} and LAMOST \citep[e.g.,][]{Deng2012, Cui2012, Liu2014, Yuan2015} surveys present a compelling opportunity for conducting such research.
It can be attributed to the significant strengths of these RC stars, including:
(i) the luminosities of RC stars are quite stable and are only slightly dependent on age and chemical composition, and therefore can be used as standard candles for the distance indicators \citep[e.g.,][]{Cannon1970, Paczynski1998};
(ii) they are widely distributed across the entire Galactic disk, and thus as excellent tracers to study the 3D structures \citep{Bovy2016}, explore the chemical, kinematic and dynamic properties \citep[e.g.,][]{Huang2016, Sun2020, Sun2023, Sun2024}, and reveal the assemblage history of the Galactic disk;
(iii) the high-precision measurements of the chemical and kinematic parameters \citep[i.e., 3D velocities, 3D positions, metallicity and abundances;][]{Bovy2014, Huang2020}.
Leveraging these samples, it is possible to make an exhaustive analysis of the chemical, kinematic, and dynamic properties of the thin and thick disks over a wider range of Galactocentric radii, thereby enhancing the constraint of the origins of the thin and thick disks.
Note that, even though splitting the Galactic disk into "thin" and "thick" components has a long precedence in the literature \citep[e.g.,][]{Gilmore1983, Fuhrmann1998, Lee2011,  Mackereth2019, Han2020, Sun2024}, it is at least worth mentioning that there are arguments against doing so in favor of a continuous disk model \citep[e.g.,][]{Bensby2007, Bovy2012a, Kawata2016, Hayden2017}.
These samples could also nicely contribute to this debate as well.

This paper is structured as follows.
In Section\,2, we describe the data used in this study.
The metallicity distributions of the thin/thick disks are presented in Section\,3, and we investigate the $\Delta$V$_{\phi}$/$\Delta$[Fe/H] and $\Delta \sigma_{\phi}$/$\Delta$[Fe/H] distributions of the thin/thick disks in Section\,4.
In Section\,5, we present the distributions of the eccentricities of the thin/thick disks, and the possible formation and evolution history of the thin/thick disks is discussed in Section\,6.
Finally, our main conclusions are summarized in Section\,7.

\section{Data}
\subsection{The LAMOST and APOGEE RC samples}
We mainly use 171,320 primary RC stars \citep{Bovy2014, Huang2020} selected from the LAMOST \citep[e.g.,][]{Deng2012, Cui2012, Liu2014, Yuan2015} and APOGEE \citep{Majewski2017} surveys.
These RC stars are selected by using a method based on the large-scale spectroscopic data \citep[e.g.,][]{Bovy2014, Huang2015, Huang2020}. 
This method first selected RC-like stars in the metallicity-dependent effective temperature ($T_{\rm eff}$)$-$surface gravity (log\,$g$) plane.
Secondly, removing the secondary RC stars from the RC-like stars in the metallicity ([$Z$])$–$color ($J$\,$-$\,$K_{s}$)$_{0}$ plane.
For the APOGEE RC stars \citep[e.g.,][]{Bovy2014}, the selection of the RC-like stars used the cuts of 1.8\,$\leq$\,log\,$g$\,$\leq$\,0.0018\,dex\,K$^{-1}$\,$\big ($\,$T_{\rm eff}$\,$-$\,$T_{\rm eff}^{\rm ref}\,\big )$\,$+$\,2.5, where $T_{\rm eff}^{\rm ref}$ = $-$382.5\,K\,dex$^{-1}$\,[Fe/H]\,$+$\,4607\,K.
Then removing the secondary RC stars from the RC-like stars using the conditions of [$Z$]\,$>$\,1.21\,[($J$\,$-$\,$K_{s}$)$_{0}$\,$-$\,0.05]$^{9}$\,$+$\,0.0011 and [$Z$]\,$<$\,2.58\,[($J$\,$-$\,$K_{s}$)$_{0}$\,$-$\,0.40]$^{3}$\,$+$\,0.0034, with additional bounds of [$Z$]\,$\leq$\,0.06 and ($J$\,$-$\,$K_{s}$)$_{0}$\,$\geq$\,0.5.
For a more detailed analysis of the selection of the APOGEE RC stars, please refer to Bovy et al. ({\color{blue}{2014}}).
For the LAMOST RC stars, Huang et al. ({\color{blue}{2015, 2020}}) developed the method in Bovy et al. ({\color{blue}{2014}}) to be suitable for the LAMOST data, the cuts of the selecting RC-like stars to 1.8\,$\leq$\,log\,$g$\,$\leq$\,0.0006\,dex\,K$^{-1}$\,$\big ($\,$T_{\rm eff}$\,$-$\,$T_{\rm eff}^{\rm ref}\,\big )$\,$+$\,2.5, where $T_{\rm eff}^{\rm ref}$ = $-$873.1\,K\,dex$^{-1}$\,[Fe/H]\,$+$\,4255\,K, with additional bounds of $T_{\rm eff}$\,$\leq$\,5000\,K and log\,$g$\,$\leq$\,2.75.
The conditions of removing the secondary RC stars from the RC-like stars update to [$Z$]\,$>$\,1.21\,[($J$\,$-$\,$K_{s}$)$_{0}$\,$-$\,0.085]$^{9}$\,$+$\,0.0011 and [$Z$]\,$<$\,2.58\,[($J$\,$-$\,$K_{s}$)$_{0}$\,$-$\,0.400]$^{3}$\,$+$\,0.0034.
For details on the selection of the LAMOST RC stars, please refer to Huang et al. ({\color{blue}{2015, 2020}}).
By selecting the sources with large signal-to-noise ratio (SNR) for the common targets of the LAMOST and APOGEE RC stars, 177,123 primary RC stars are further selected, 137,448 primary RC stars from LAMOST and 39,675 primary RC stars from APOGEE.

The stellar parameters, such as, $T_{\rm eff}$, log\,$g$, metallicity ([Fe/H]) and [$\alpha$/Fe], of the APOGEE and LAMOST data, are determined by the ASPCAP \citep[e.g.,][]{Garcia2016} and LSP3 pipelines \citep[e.g.,][]{Xiang2015}, respectively.
With various tests, the typical uncertainties of the $T_{\rm eff}$, log\,$g$, [Fe/H] and [$\alpha$/Fe] for the LAMOST sample stars, are respectively better than around, 100\,K, 0.10\,dex, 0.10$-$0.15\,dex and 0.03$-$0.05\,dex, for the APOGEE sample stars, the uncertainties are slightly smaller \citep[see][]{Bovy2014, Xiang2015, Garcia2016, Huang2020}.
Due to the standard candle nature of the RC stars, the distance errors are around 5\%-10\%.
To ensure the accuracy of the kinematic and dynamic calculations,  the astrometric parameters (e.g., proper motions) of the sample stars are updated by using the Gaia DR3 catalog \citep[e.g.,][]{Gaia Collaboration2023a, Gaia Collaboration2023b, Recio-Blanco2023}.

\begin{figure}[t]
\begin{center}
\includegraphics[width=8.5cm]{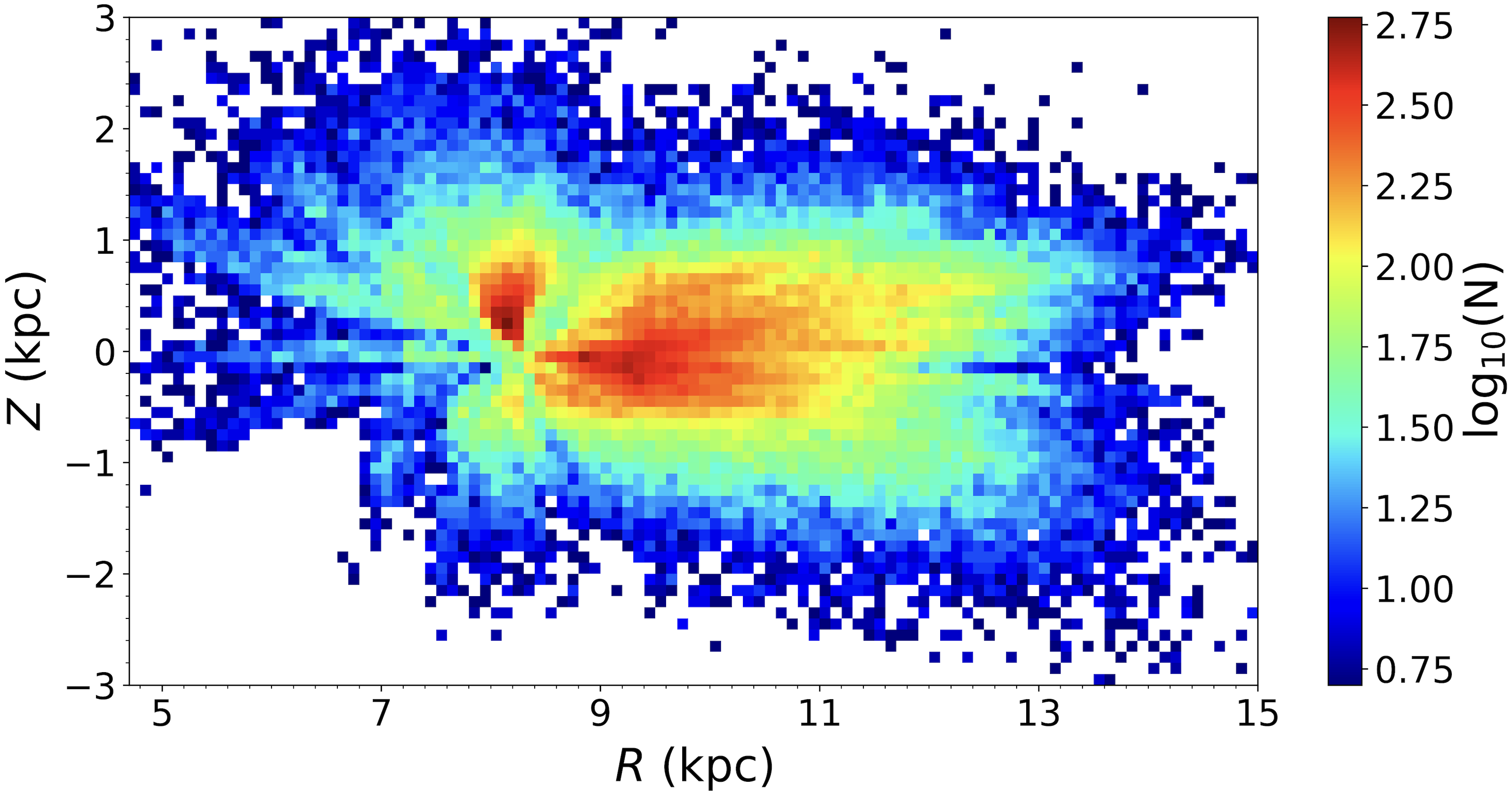}
\caption{Spatial distribution in the $R$ - $Z$ plane, of the sample stars, with color-coded by the stellar number density.
There is a minimum of 8 stars per bin spaced 0.1\,kpc in both axes.}
\end{center}
\end{figure}
%%\label{fig1}

\begin{figure*}[t]
\centering
\subfigure{
\includegraphics[width=8.6cm]{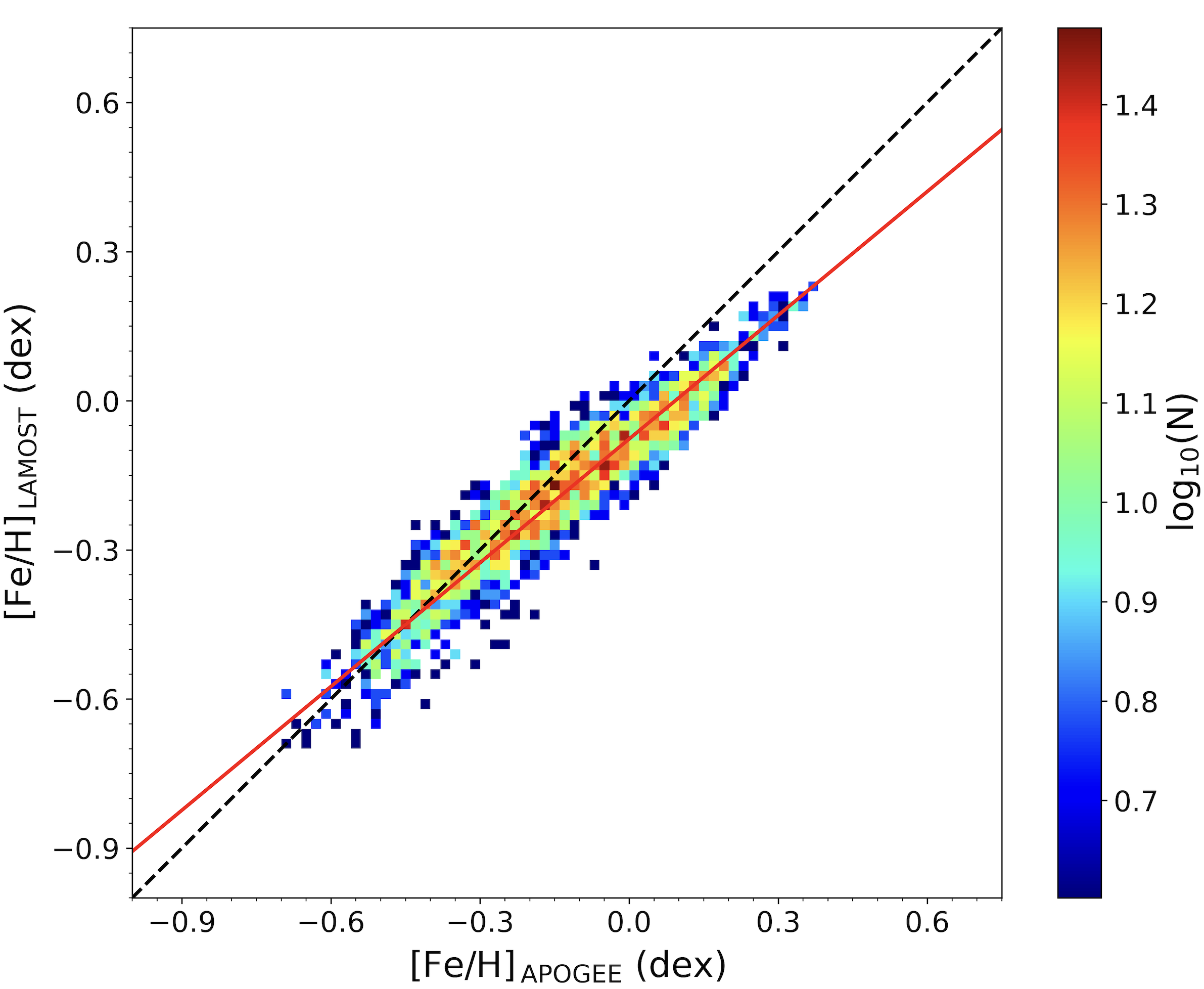}
}
\hspace{0.25cm}
\subfigure{
\includegraphics[width=8.6cm]{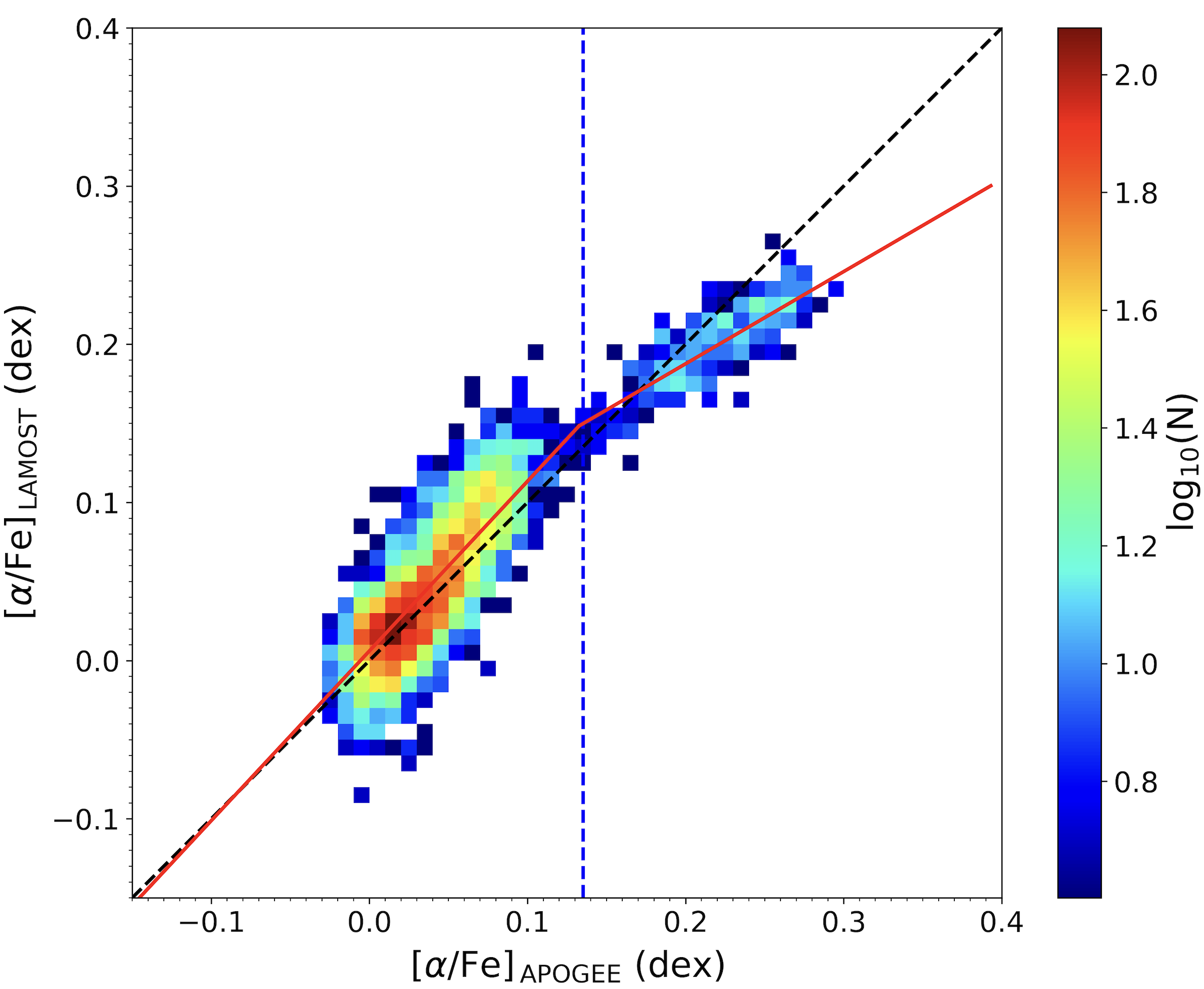}
}

\begin{center}
\caption{The [Fe/H] (left panel) and [$\alpha$/Fe] (right panel) calibrations between the APOGEE and LAMOST datasets.
Black dashed lines represent the one-to-one relations, and the red solid lines are linear fits to the data.
There is a minimum of 4 stars per bin, with the [Fe/H] and [$\alpha$/Fe] respectively spaced in 0.02\,dex and 0.02\,dex.}
\end{center}
\end{figure*}
%%\label{fig2}

\subsection{Coordinate Systems and Sample Selections}

The standard Galactocentric cylindrical coordinates ($R$,\,$\phi$,\,$Z$) are adopted, and the 3 velocity components are $V_{R}$, $V_{\phi}$ and $V_{Z}$, respectively.
To estimate the kinematics and orbits, the Galactocentric distance of the Sun, the local circular velocity and the solar motions are set to be $R_{0}$ = 8.34 kpc \citep{Reid2014}, $V_{c, R_{0}}$ = 238.0 km s$^{-1}$ \citep[e.g.,][]{Reid2004, Schonrich2010, Schonrich2012, Reid2014, Huang2015, Huang2016, Bland-Hawthorn2016}, and ($U_{\odot}$, $V_{\odot}$, $W_{\odot}$) $=$ $(13.00, 12.24, 7.24)$ km s$^{-1}$ \citep{Schonrich2018}, respectively.
The orbital parameters (i.e. the eccentricity) of each star are determined by {\it Galpy} \citep{Bovy2015}, under the Galactic gravitational potential as ``$MWPotential2014$".
To further ensure the accuracy of the kinematics and dynamics, the following cuts are used for the sample stars:

\begin{itemize}[leftmargin=*]

\item  {LAMOST and APOGEE spectral SNRs $\geq$ 20;}

\item  {Distance error $\leq$ 10\%;}

\item  {[Fe/H] $\geq -1.0$ dex and $|V_{Z}|$ $\leq$ 120 km s$^{-1}$.}

\end{itemize}
The first two cuts ensure the uncertainties of the 3D velocities of the sample stars are generally within 5.0\,km\,s$^{-1}$.
The last cut is used to exclude the halo stars \citep[e.g.,][]{Huang2018, Hayden2020, Sun2020}.
By the above cuts, 170,729 RC stars are finally selected, of which 39,112 and 131,617 stars are respectively from the APOGEE and LAMOST survey.
The spatial distribution of the final selected stars is shown in Fig.\,1.
Since the [Fe/H] and [$\alpha$/Fe] are typically different for the two datasets, considering that the LAMOST has a larger sample size than the APOGEE, we further calibrate these parameters of the APOGEE dataset to the LAMOST system based on the best linear fit as displayed in Fig.\,2.

\begin{table*}
\caption{The properties of the thin/thick disks}

\centering
\setlength{\tabcolsep}{6mm}{
\begin{tabular}{lllllllll}
\hline
\hline
\specialrule{0em}{5pt}{0pt}
Name                                         & $\langle$ [$\alpha$/Fe] $\rangle$  &  $\langle$ [Fe/H] $\rangle$  &     $\left\langle R \right\rangle$   &          $\left\langle V_{\phi} \right\rangle$        &  $\sigma_{\phi}$  &     Number     &    Fraction \\
                                             &                (dex)               &        (dex)       &     (kpc)     &        (km\,s$^{-1}$)           &    (km\,s$^{-1}$)    &           &               \\
\specialrule{0em}{5pt}{0pt}
\hline
\specialrule{0em}{3pt}{0pt}
Thin disk   &  0.03   &            $-$0.2             &     10.14     &            225.01               &          22.97       &    135,009 &      79.08\%\\ [0.2cm]
Thick disk  &  0.20   &            $-$0.4            &     8.54      &            179.27               &          54.81       &   23,168  &      13.57\%   \\
\specialrule{0em}{3pt}{0pt}
\hline
\specialrule{0em}{3pt}{0pt}
\end{tabular}}
\label{tab:datasets}
\end{table*}

\subsection{The thin and thick disks of the RC sample}
Since the spatial distributions and kinematics of stars can be modified over time, while the atmospheric chemical abundance of stars is generally invariant, we separate the thin and thick disks by using [$\alpha$/Fe] ratio.
The [Fe/H]--[$\alpha$/Fe] relation of our sample stars is shown in Fig.\,3.
The plot shows clearly two branches, which are respectively the thin and thick disks.
To get pure thin and thick disks, inspired from previous work  \citep[e.g.,][]{Bensby2005, Lee2011, Brook2012, Haywood2013, Recio-Blanco2014, Nidever2014, Guiglion2015, Hayden2015, Queiroz2020, Sun2023, Sun2024}, two empirical cuts (see Fig.\,3) are applied to select the thick disk (23,168 stars above the upper cut) and thin disk (135,009 stars below the lower cut), and the properties of these two disks are presented in details in Table 1.

The histograms of the fractional number density of the distributions for vertical and radial directions, metallicity and $V_{\phi}$ are shown in Fig.\,4. The results indicate that the thin disk stars are mainly distributed within $|Z| < 3.0$\,kpc, while the thick disk stars have a relatively extended distribution with the stellar height extending to up to $|Z| \sim$ 5.0\,kpc.
This is in rough agreement with the nature of the scale-height of the thin disk \citep[typically suggested as $h_{z}$ = 0.3$-$0.6\,kpc; e.g.,][]{Bovy2012b, Bland2018} is smaller than that of the thick disk \citep[typically suggested as $h_{z}$ = 0.8$-$1.0\,kpc; e.g.,][]{Bovy2012b, Bland2018}.
For the radial direction, the thick disk stars peak around 8.0\,kpc, while the thin disk stars peak around a much larger Galactic radius, that is, $R \sim 9.5$\,kpc, which is in line with the scale-length of the thin disk \citep[typically suggested as 2.5$-$3.6\,kpc; e.g.,][]{Bovy2012b, Hayden2015} is larger than that of the thick disk \citep[typically suggested as 1.8$-$2.2\,kpc; e.g.,][]{Bensby2011, Bovy2012b, Hayden2015}.
The [Fe/H] of the thin and thick disks respectively peaks around $-$0.2\,dex and $-$0.5\,dex, which is consistent with previous studies \citep[e.g.,][]{Lee2011, Haywood2013, Sun2020, Sun2023}.
The orbital velocity of the thin disk ($V_{\phi}$ $\sim$ 230\,km\,s$^{-1}$) is larger than the thick disk ($V_{\phi}$ $\sim$ 195\,km\,s$^{-1}$), which is also in rough agreement with recent results \citep[e.g.,][]{Lee2011, Sun2024}.
This behavior indicates that the asymmetric drift of thick disk stars is generally stronger than thin disk stars.
It is worth noting that we do not correct the selection effect for the analyses in Fig.\,4, and some features caused by the selection effects are visible, as an example, the double-peaked radial distribution of the thin disk stars (see upper right of Fig.\,4).

\begin{figure}[t]
\begin{center}
\includegraphics[width=8.4cm]{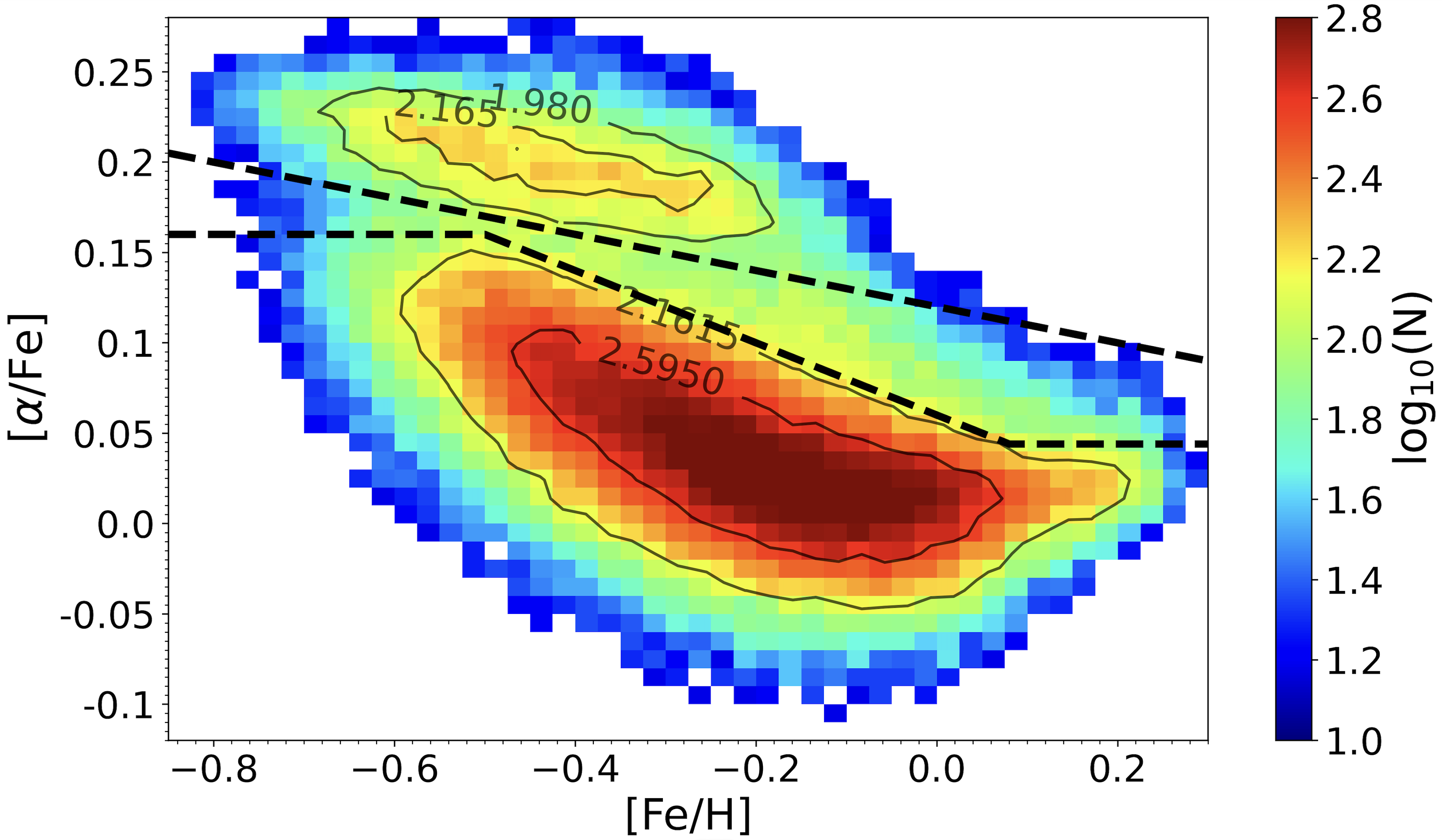}
\caption{The [Fe/H]$-$[$\alpha$/Fe] distribution of the final sample stars, overplotted with contours of equal densities, color-coded by the logarithmic number density.
The horizontal axis and vertical axis are respectively spaced by 0.025\,dex and 0.02\,dex, with a minimum of 15 stars per bin.
The two dashed lines are used to separate the thick (above the lines) and the thin (below the lines) disk stars.}
\end{center}
\end{figure}
%%\label{fig3}

\begin{figure*}[t]
\centering
\subfigure{
\includegraphics[width=8.4cm]{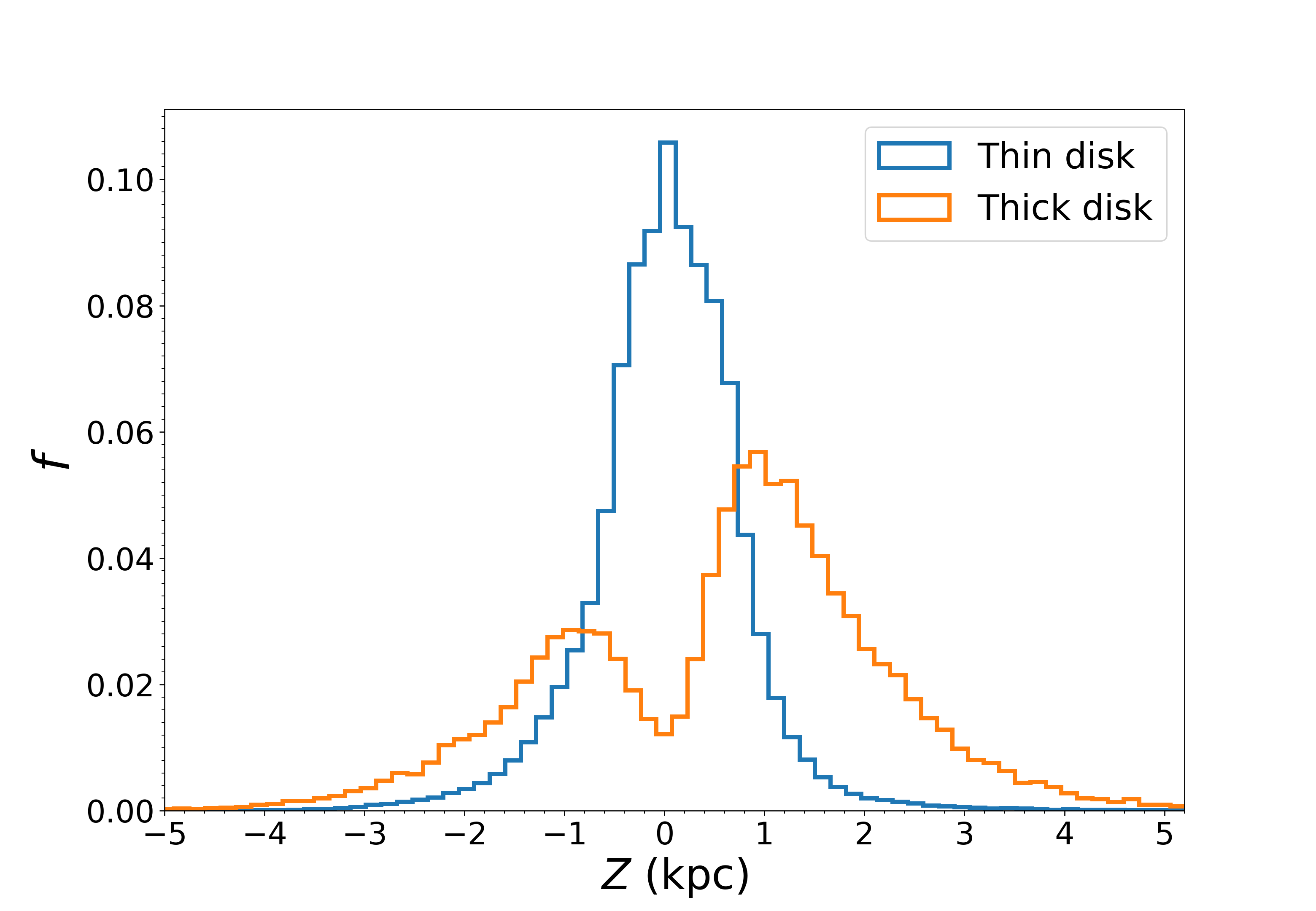}
}
\hspace{0.1cm}
\subfigure{
\includegraphics[width=8.4cm]{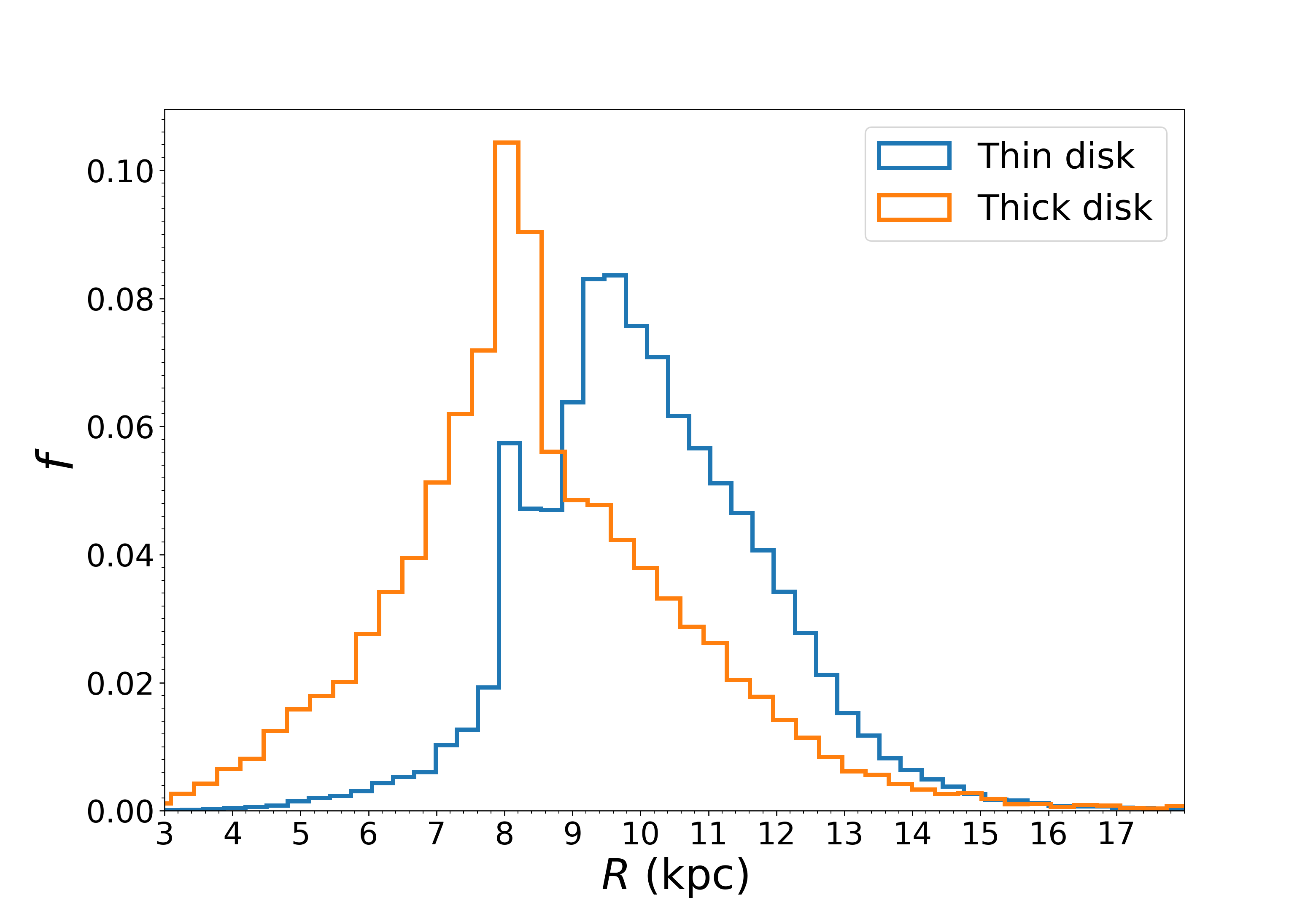}
}

\subfigure{
\includegraphics[width=8.4cm]{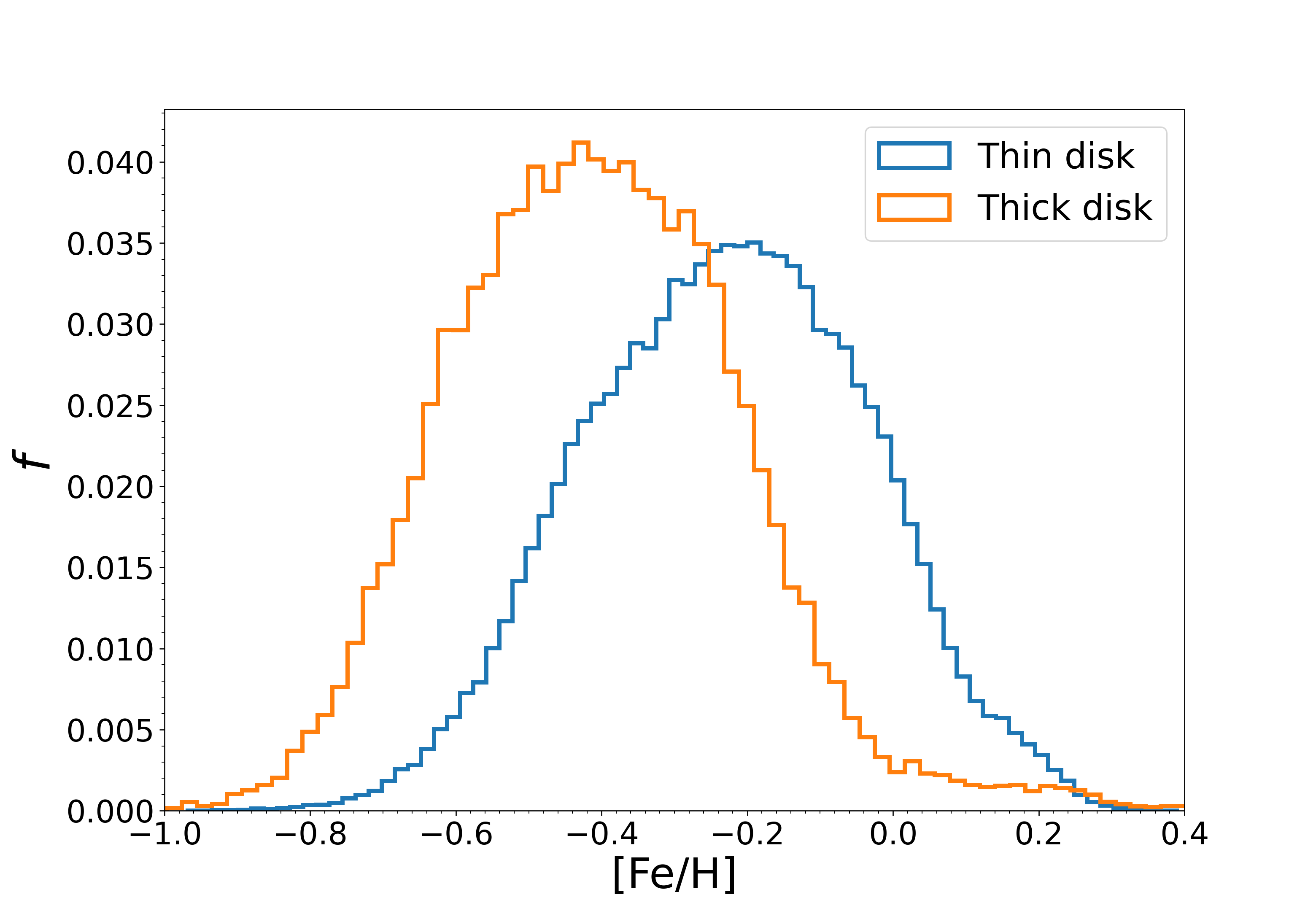}
}
\hspace{0.1cm}
\subfigure{
\includegraphics[width=8.0cm]{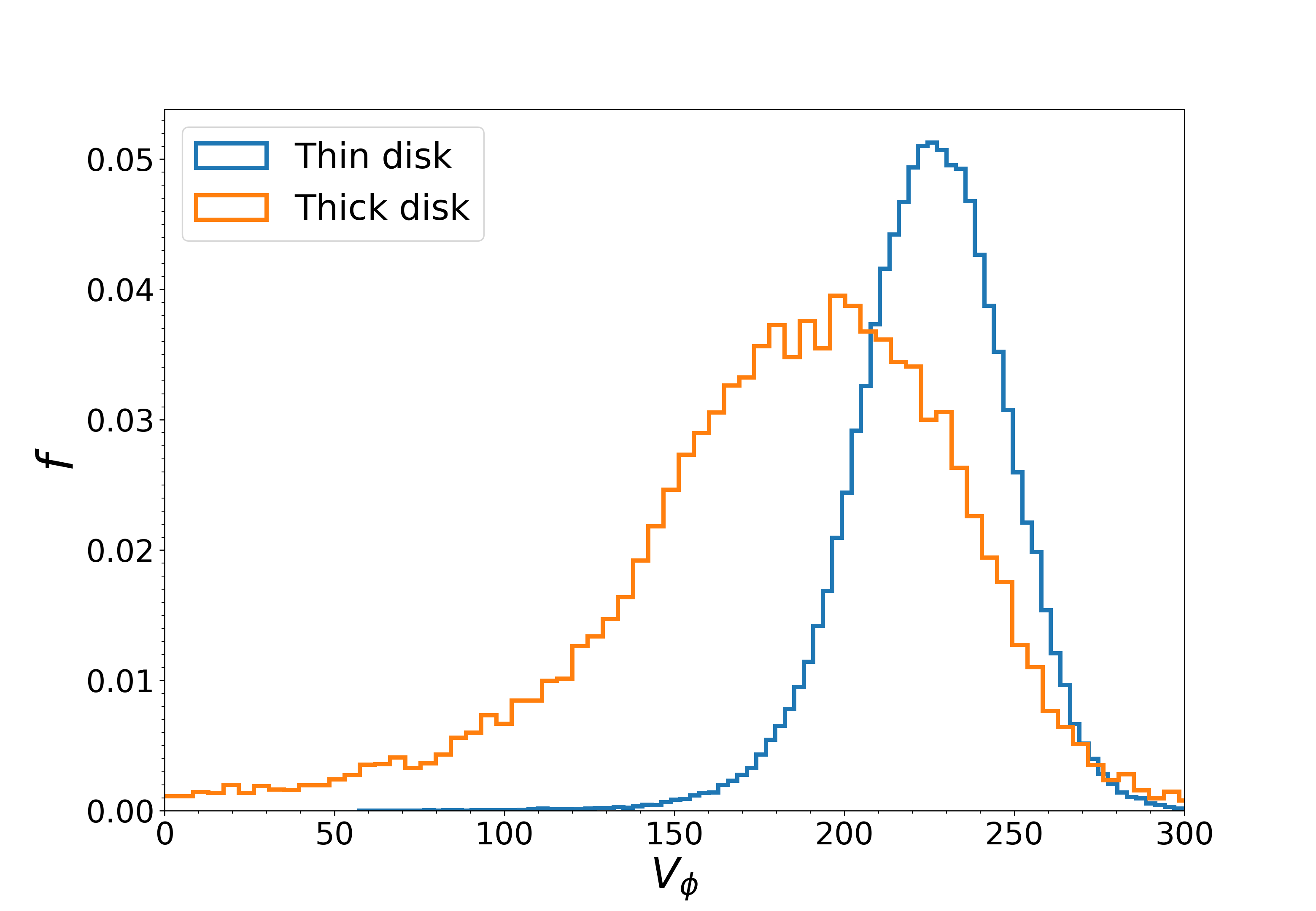}
}

\caption{Fractional number density ($f = N_i/N_{\rm tot}$) distributions of the vertical direction (upper left), radial direction (upper right), [Fe/H] (bottom left) and $V_{\phi}$ (bottom right), for the thin and thick disks.
In which, $N_i$ is the number of stars in the $i$-th radial/vertical bin and $N_{\rm tot}$ is the total number of stars of the individual population.
Different color lines represent different populations that are marked as labels in the top corner.}
\end{figure*}
%%\label{fig4}

\begin{figure*}[t]
\centering
\subfigure{
\includegraphics[width=8.4cm]{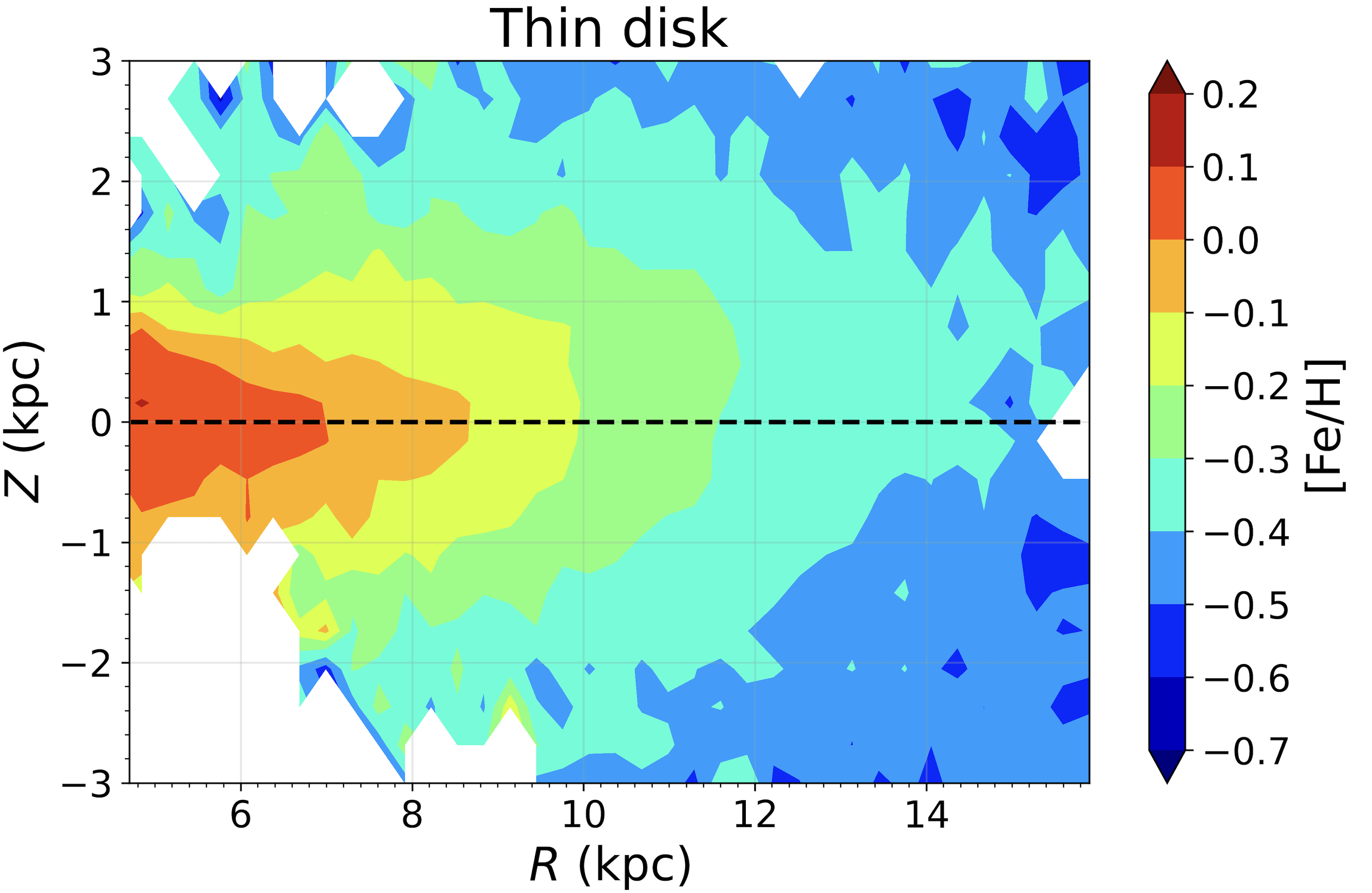}
}
\hspace{0.05cm}
\subfigure{
\includegraphics[width=8.4cm]{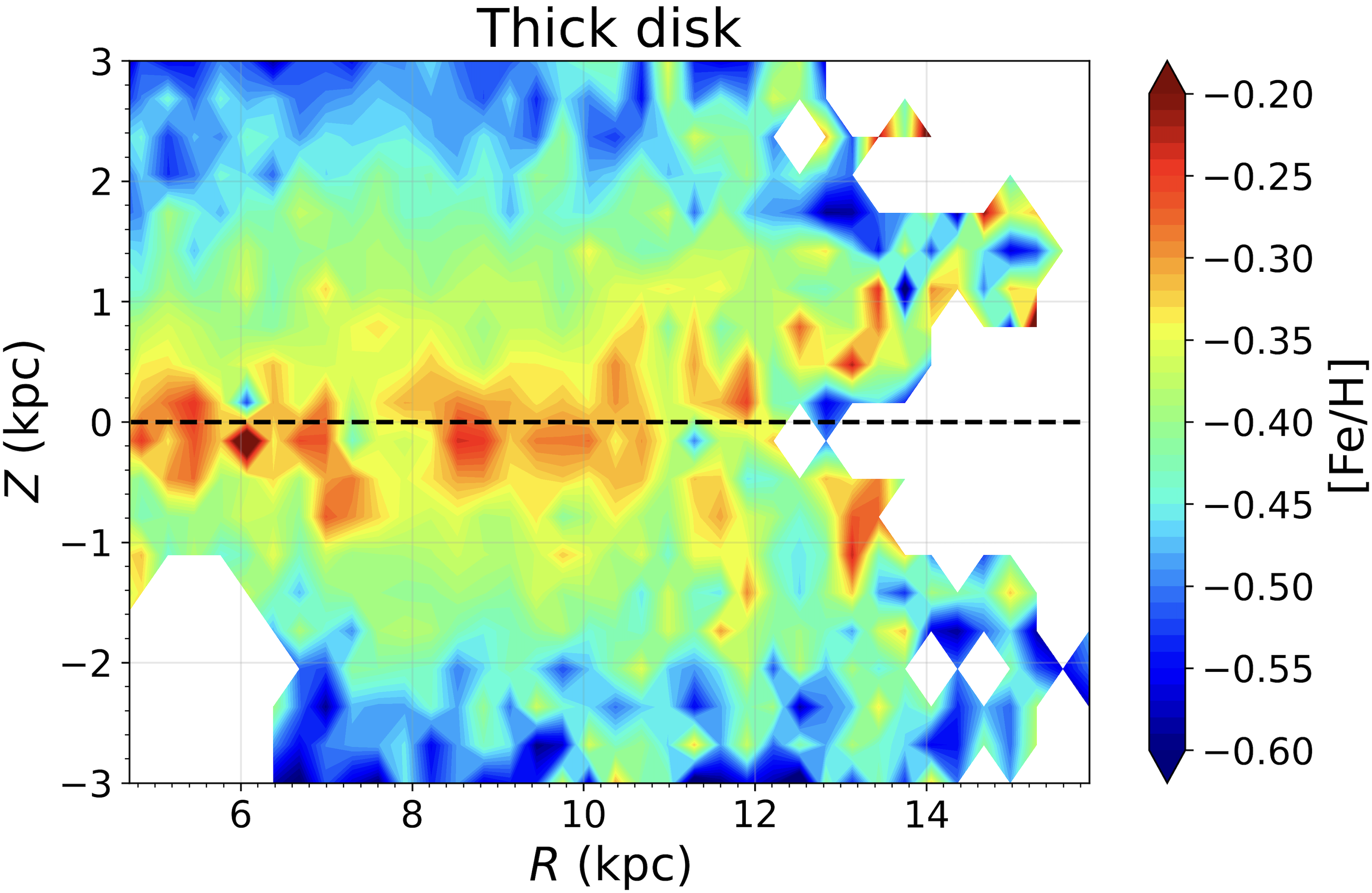}
}

\caption{Metallicity distribution, in the $R$ - $Z$ plane, of the thin (left panel) and thick (right panel) disks.}
\end{figure*}
%%\label{fig5}

\section{Metallicity distribution of the Thin and Thick Disks}

\begin{figure*}[t]
\centering
\subfigure{
\includegraphics[width=8.78cm]{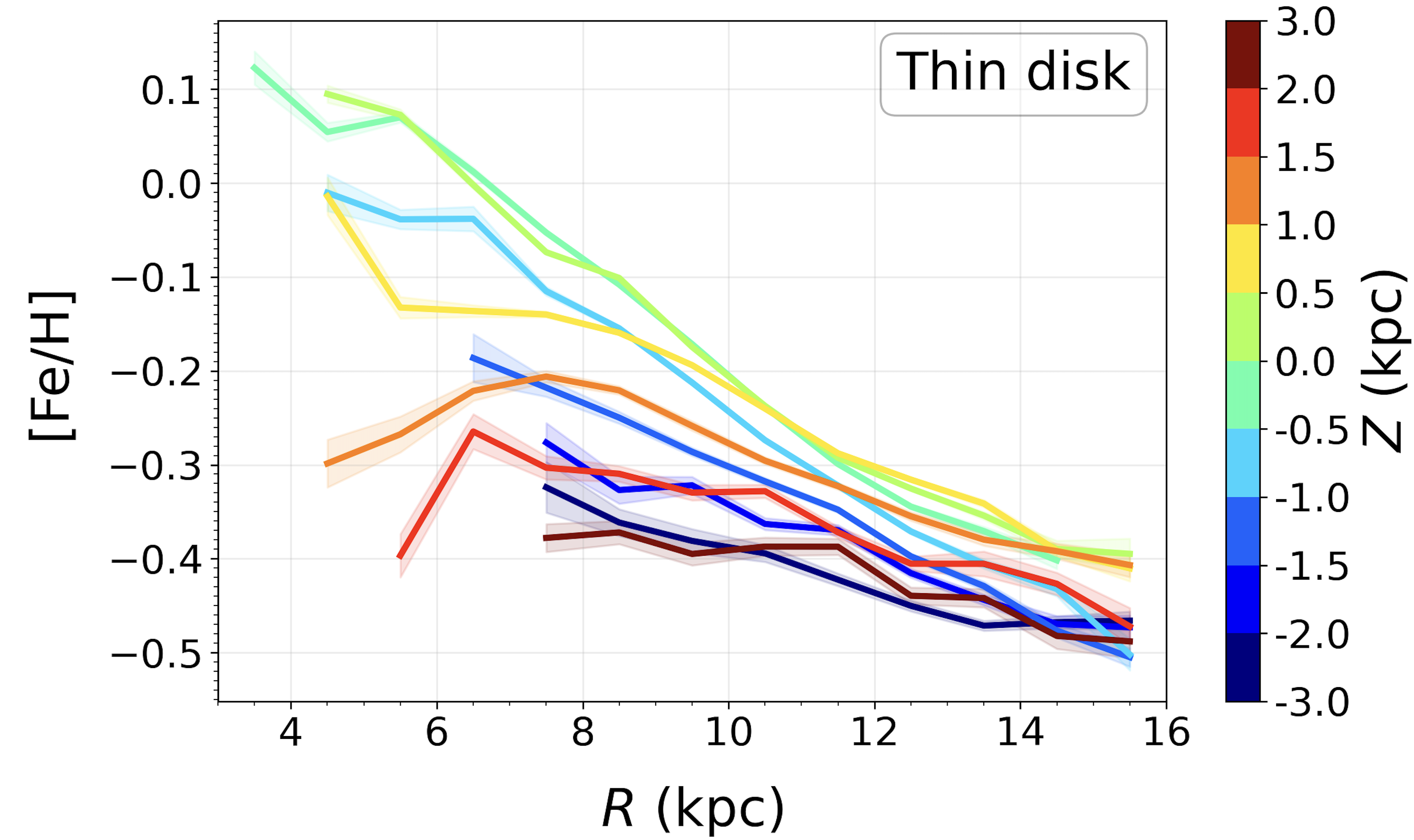}
}
\hspace{0.05cm}
\subfigure{
\includegraphics[width=8.78cm]{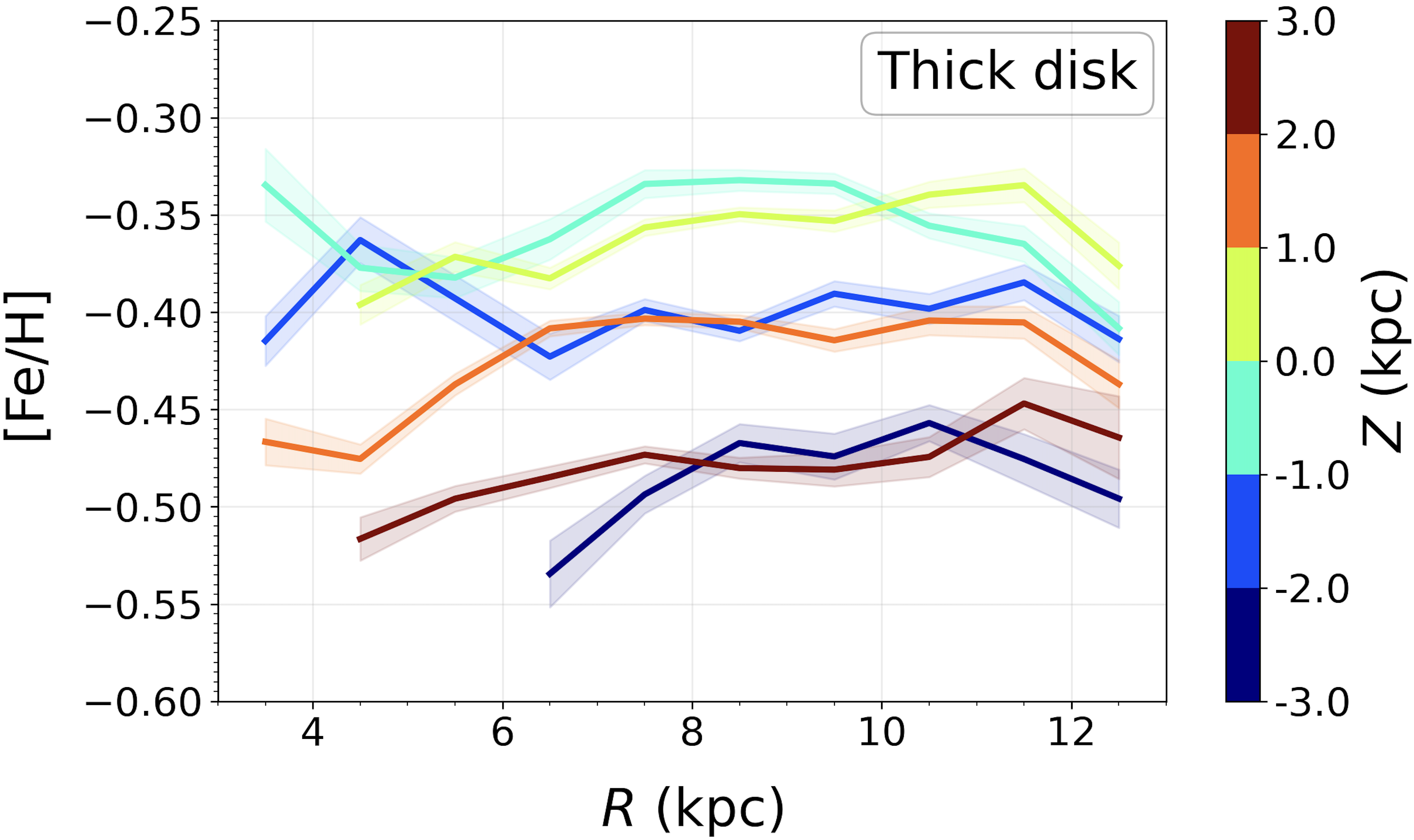}
}

\subfigure{
\includegraphics[width=8.78cm]{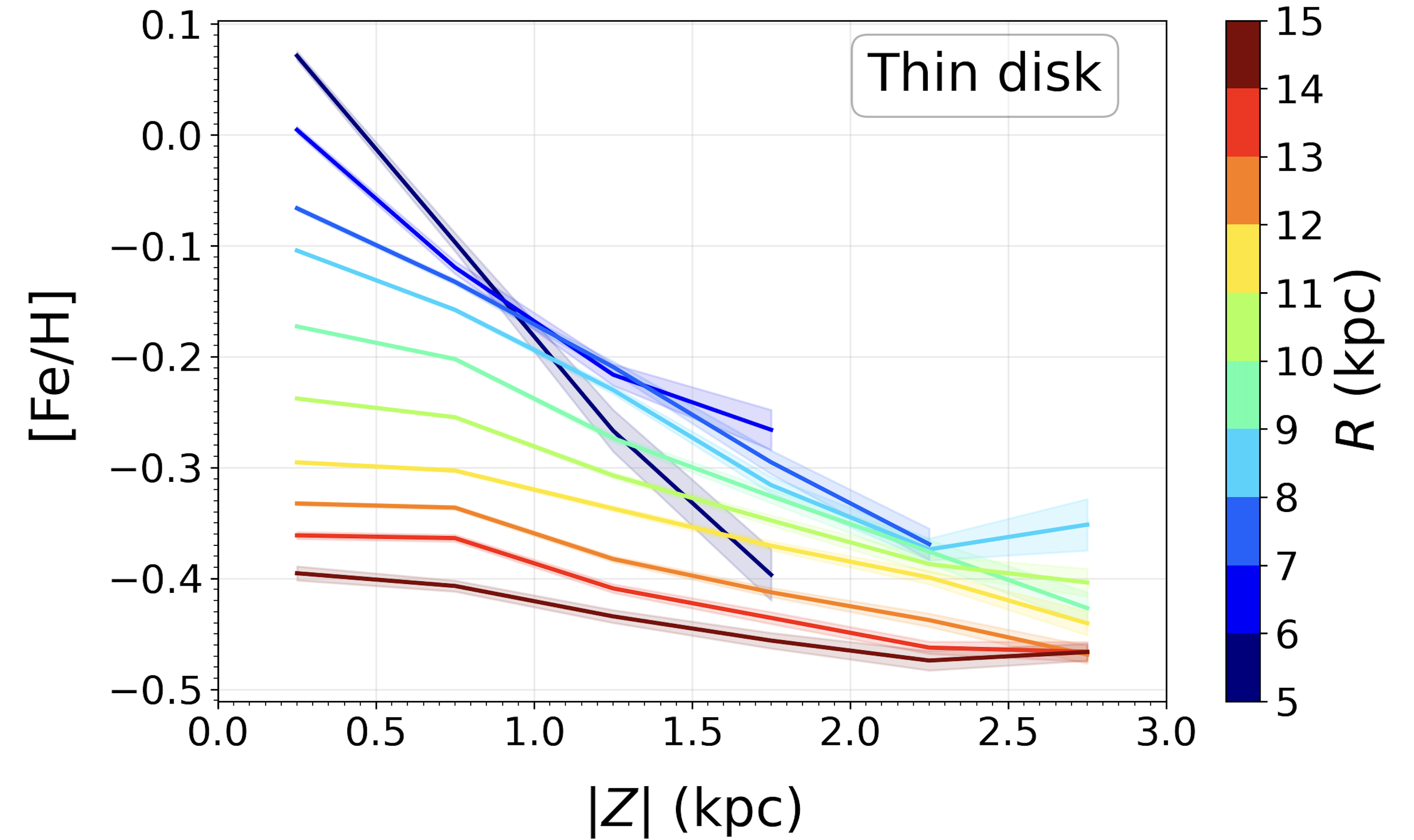}
}
\hspace{0.05cm}
\subfigure{
\includegraphics[width=8.78cm]{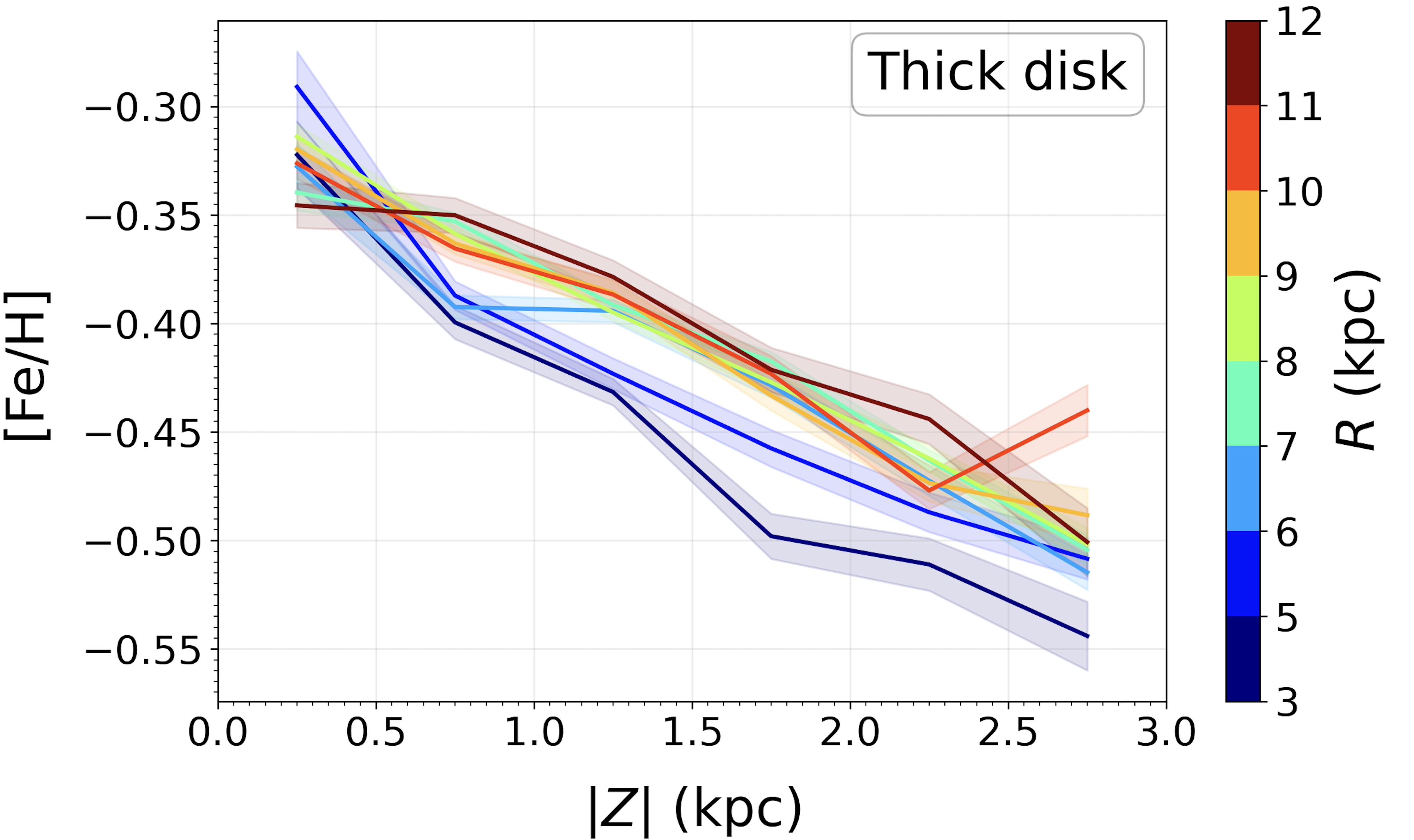}
}
\caption{Radial (upper panels) and vertical (bottom panels) metallicity profiles of the thin (left panels) and thick (right panels) disks.}
\end{figure*}
%label{fig6}

\begin{figure*}[t]
\centering
\subfigure{
\includegraphics[width=8.4cm]{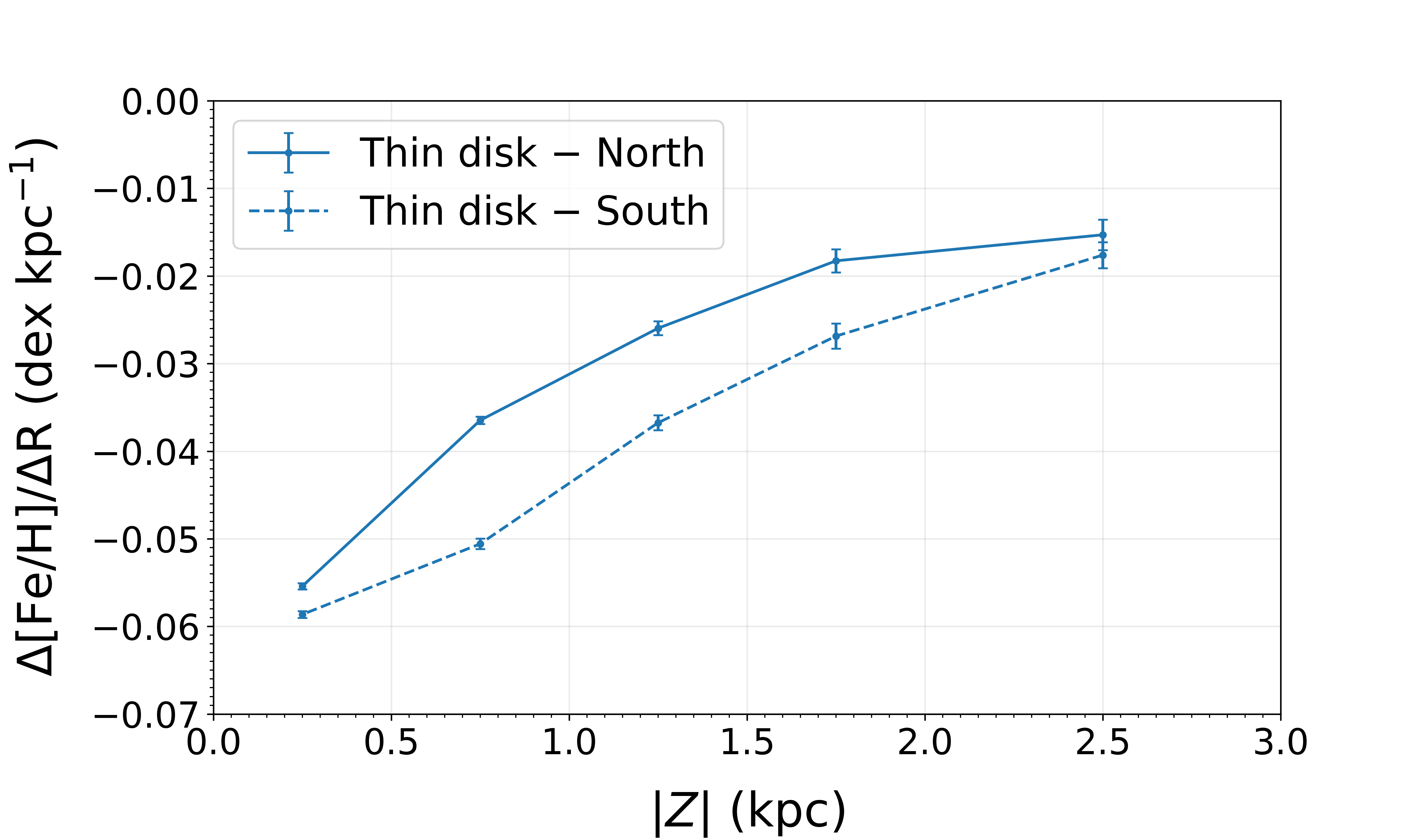}
}
\hspace{0.5cm}
\subfigure{
\includegraphics[width=8.4cm]{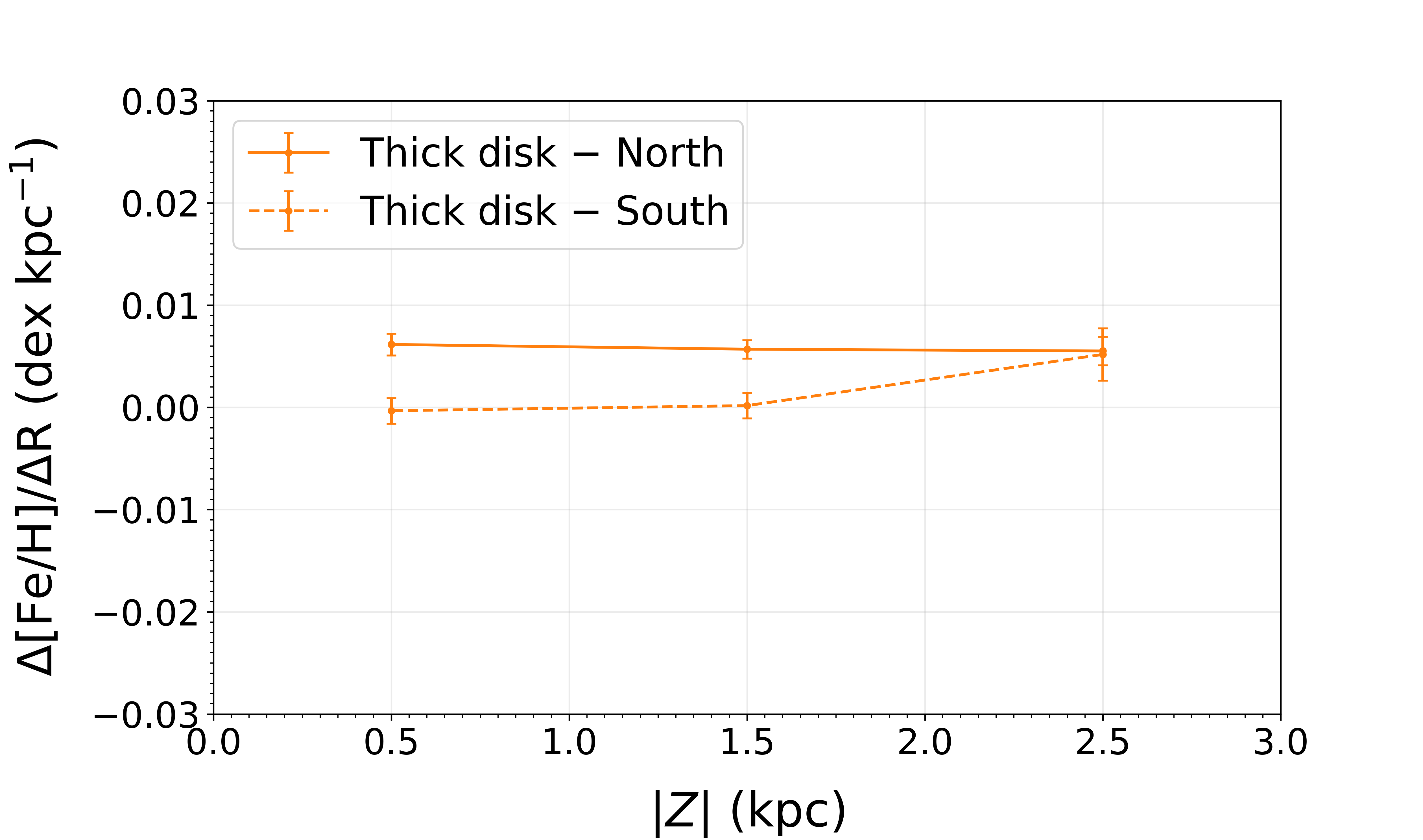}
}

\subfigure{
\includegraphics[width=8.4cm]{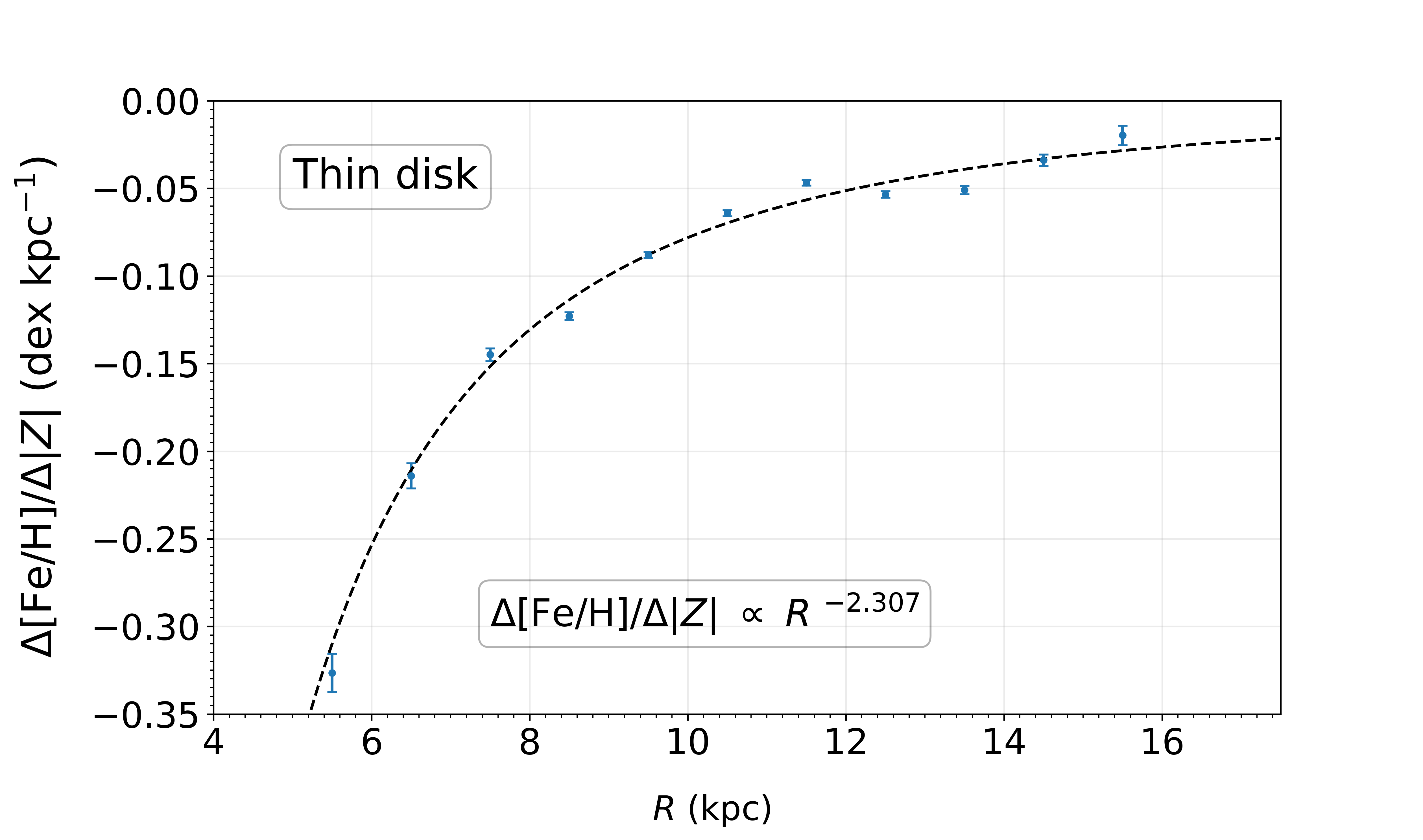}
}
\hspace{0.5cm}
\subfigure{
\includegraphics[width=8.4cm]{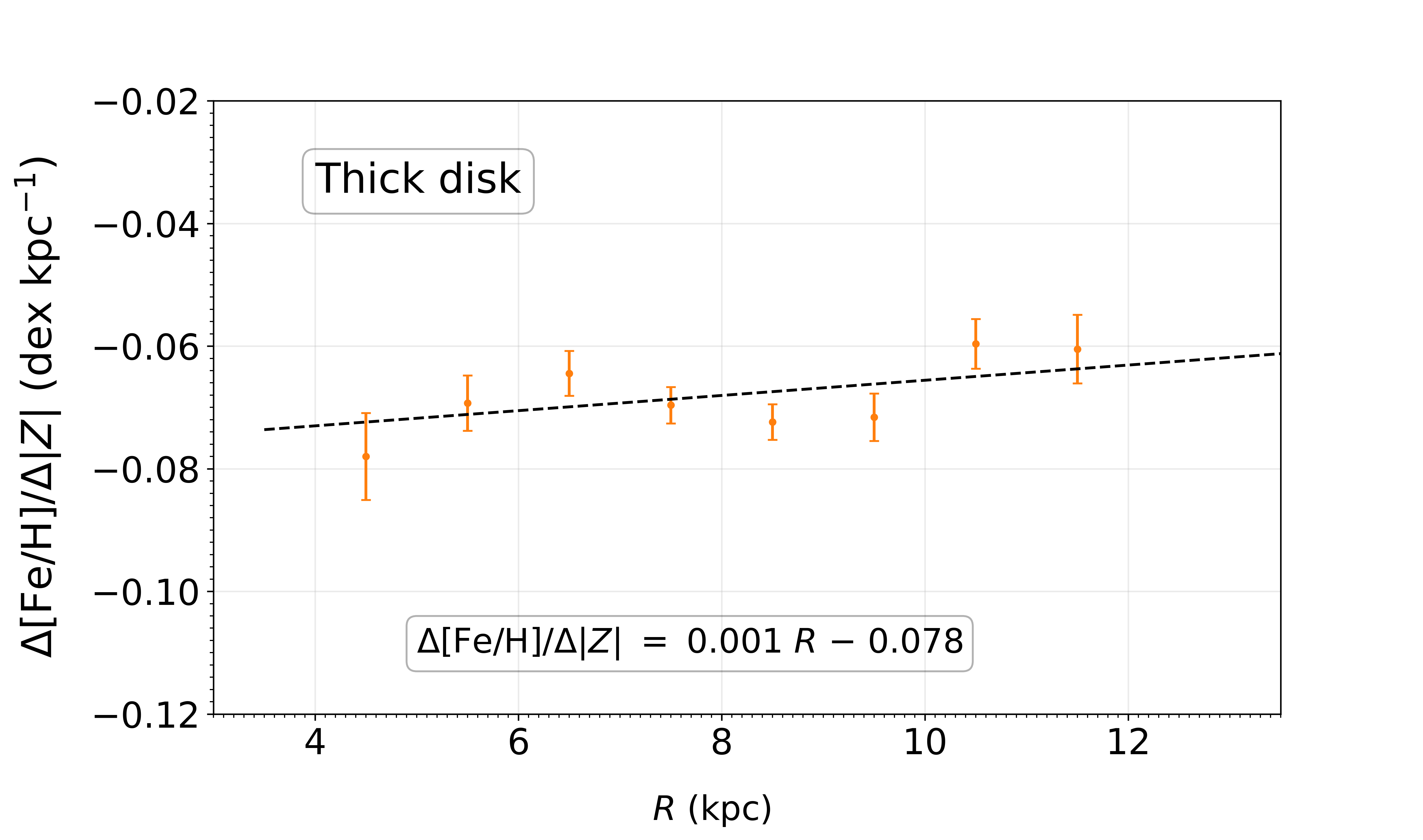}
}

\caption{The $\Delta$[Fe/H]/$\Delta$R as a function of the $|Z|$ of the thin (upper left) and thick (upper right) disks, and the $\Delta$[Fe/H]/$\Delta |Z|$ as a function of the $R$ of thin (bottom left) and thick (bottom right) disks.}
\end{figure*}
%%\label{fig7}

\begin{figure*}
\centering
\subfigure{
\includegraphics[width=8.4cm]{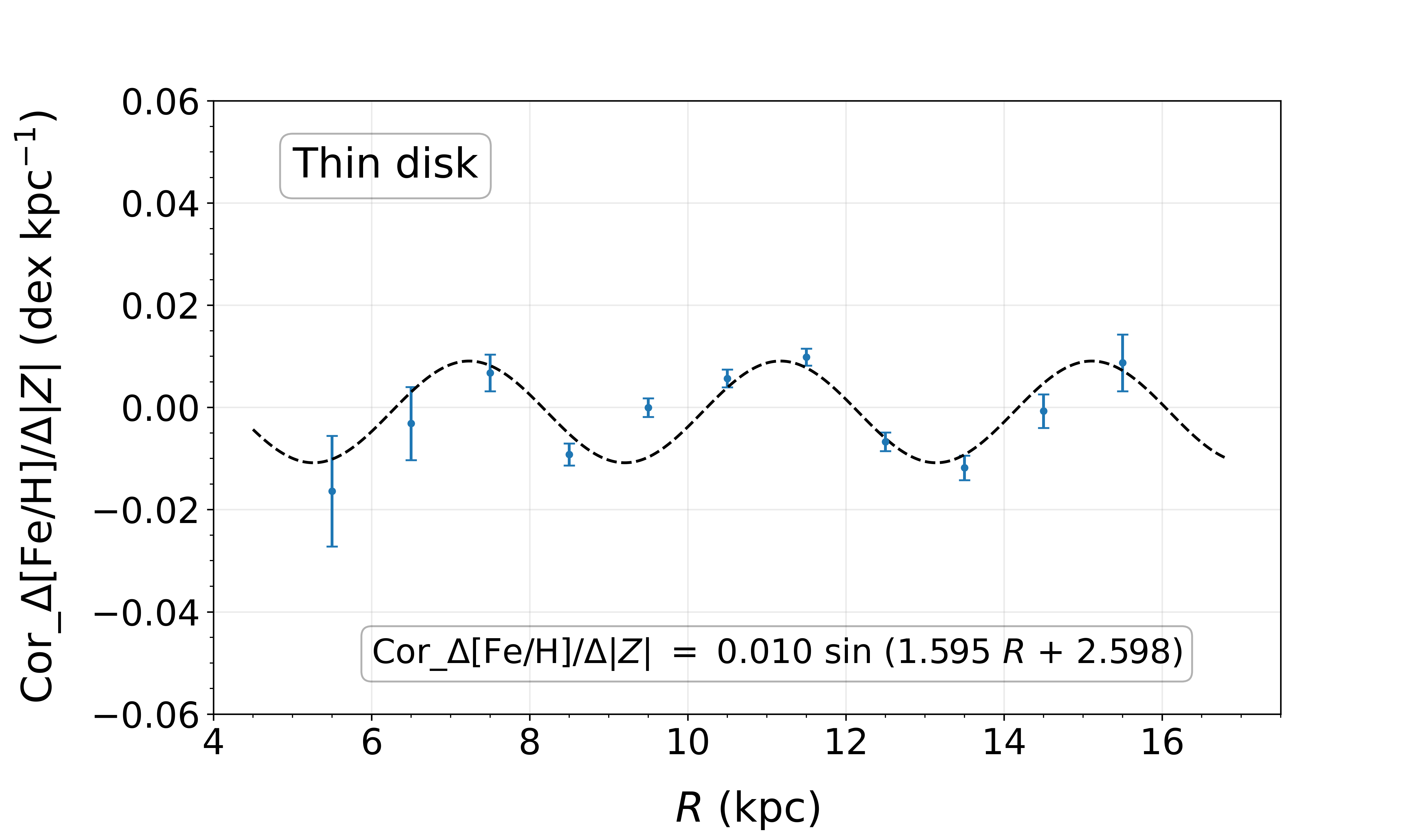}
}
\hspace{0.5cm}
\subfigure{
\includegraphics[width=8.4cm]{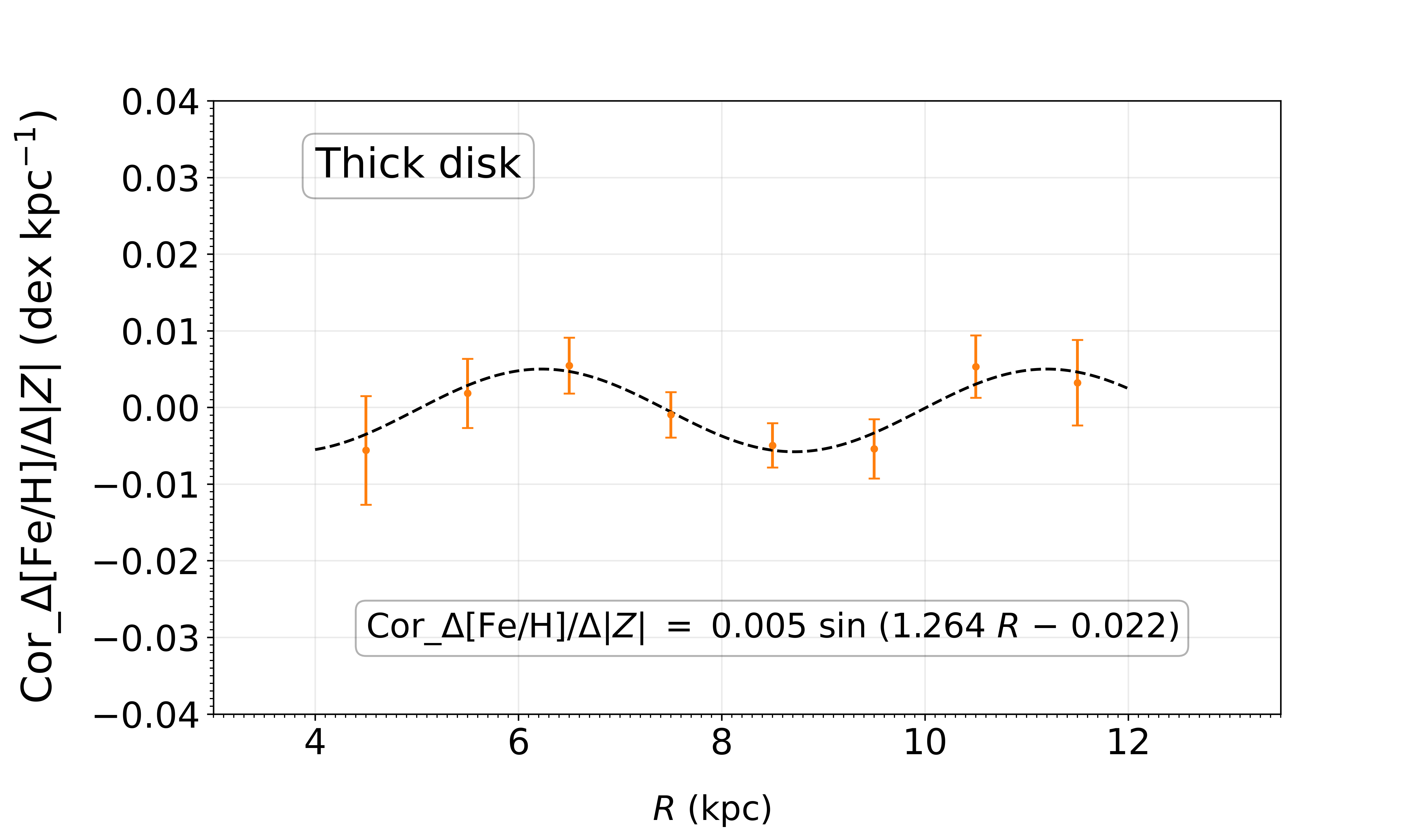}
}
\caption{The correction vertical metallicity gradient as a function of $R$ for the thin (left panel) and thick (right panel) disks.}
\end{figure*}
%%\label{fig8}\\

The results of the metallicity distributions of thin/thick disks are shown in Figs.\,5--7.
The thin disk shows obvious radial and vertical trends in [Fe/H], which decreases with $R$ (or $|Z|$).
The thick disk shows a weak decreasing trend in [Fe/H] with $|Z|$, while has no obvious trend with $R$ (see Fig.\,5).
The [Fe/H]--$R$ profile of the thin disk (see Fig.\,6) shows significant differences for various $Z$ layers, with the trend becoming flat as $R$ increases.
The global profiles break at $R$ around 8.5\,kpc, especially for $Z$ = [1.0,\,1.5]\,kpc (orange curve) and  $Z$ = [1.5,\,2.0]\,kpc (red curve), the [Fe/H] shows an increasing/flat trend with $R$.
Although the number of stars in this region is small, the limitation of the number of stars in each bin (no less than 30 stars) still ensures the statistical reliability.
Therefore, the contamination by the thick disk stars may be the possible reason for the $R_{\text{break}}$ $\sim$ 8.5\,kpc since the flat or inverse/increasing trends at $R$ $\geq$ 8.5\,kpc (see Fig.\,6).
Previous studies also suggested the radial metallicity profile breaks at the outer disk \citep[e.g.,][]{Spina2021, Yong2012, Donor2020}, this break in this work is also traced with $R_{\text{break}}$ $\sim$ 13.0\,kpc (see Fig.\,6) and is in good agreement with Spina et al. ({\color{blue}{2021}}).
Since the cut-off radius of the disk star formation is generally suggested not to be larger than $\sim$ 14.0 kpc \citep{Schonrich2017},  the flared migrators may be the possible reason for this break.
The profiles of [Fe/H]--$|Z|$ of the thin disk show a linear function shape, with the trend becoming flat as $R$ increases (see Fig.\,6).

%%%%%%%%%%%%
%%%%%%%%%%%
%%%%%%%%%%%%

The results of $\Delta$[Fe/H]/$\Delta$R as function of $|Z|$ and the $\Delta$[Fe/H]/$\Delta|Z|$ as function of $R$, of the thin/thick disks, are shown in Fig.\,7.
For $\Delta$[Fe/H]/$\Delta$R, the thin disk shows an increasing trend with $|Z|$, rising from around $-$0.06\,dex\,kpc$^{-1}$ at $|Z|$ $<$ 0.25\,kpc to $-$0.02\,dex\,kpc$^{-1}$ at $|Z|$ $>$ 2.75\,kpc.
The thick disk is generally weaker than $\sim$ 0.01 dex\,kpc$^{-1}$, and shows no obvious change with $|Z|$, which is in rough agreement with the result of old/high-$\alpha$ populations \citep[e.g.,][]{Imig2023}.
This may be understood because strong migration can wipe out any possible original gradients.
However, the positive $\Delta$[Fe/H]/$\Delta$R of the thick disk \citep[this positive gradient has been also observed in other works, e.g.,][]{Vickers2021, Imig2023} is surprising, it means that we do seem to see some evidence of the inside-out star formation and the stars are not fully mixed up to now.
In addition, the $\Delta$[Fe/H]/$\Delta$R of the thin disk is obviously stronger in the south disk than the north disk.

Using a suite of cosmological chemo-dynamical disk galaxy simulations, Miranda et al. ({\color{blue}{2016}}) reported $\Delta$[Fe/H]/$\Delta$R as a function of $|Z|$ for thin and thick disks of the Milky Way-like galaxies simulated in a cosmological context, at 5.0 $<$ $R$ $<$ 10.0\,kpc.
For thin disk, Miranda et al. ({\color{blue}{2016}}) suggested the thin disk in the region of $|Z|$ = 1.0$-$1.5\,kpc has $\Delta$[Fe/H]/$\Delta$R = $-$0.028\,dex\,kpc$^{-1}$.
This result is in good agreement with ours for the thin disk in the north with $|Z|$ = 1.0$-$1.5\,kpc (see upper left panel of Fig.\,7).
Their thin disk result outside the region of $|Z|$ = 1.0$-$1.5\,kpc is obviously different from ours, which may be caused by the condition with 5.0 $<$ $R$ $<$ 10.0\,kpc of Miranda et al. ({\color{blue}{2016}}).
It is worth mentioning that our thin disk results in the region of $|Z|$ = 0.0$-$0.5 are in rough agreement with Pilkington et al. ({\color{blue}{2012}}), who reported the young population of the galaxies selected from the RaDES (Ramses Disk Environment Study) has $\Delta$[Fe/H]/$\Delta$R = $-$0.059\,dex\,kpc$^{-1}$.
For thick disk, the results with $|Z|$ $<$ 1.5\,kpc of Miranda et al. ({\color{blue}{2016}}), are in good agreement with ours, while their gradients at $|Z|$ $>$ 1.5\,kpc are obviously stronger than our results (see upper right panel of Fig.\,7), while it is in good agreement with the metallicity gradient in guiding radius ($R_{g}$) of the thick disk in Vickers et al. ({\color{blue}{2021}}).
These differences may be related to the stellar migration of the thick disk stars \citep[e.g.,][]{Sellwood2002, Han2020, Vickers2021}.

For $\Delta$[Fe/H]/$\Delta|Z|$, the thin disk shows an increasing trend with $R$ in the form of a single power law, which increases steadily from around $-$0.36\,dex\,kpc\,$^{-1}$ at $R$ $\sim$ 5.5\,kpc to around $-$0.05\,dex\,kpc\,$^{-1}$ at $R$ $>$ 11.5\,kpc.
The thick disk displays a similar gradient (nearly $-$0.06 $\sim$ $-$0.08\,dex\,kpc\,$^{-1}$) for all $R$ bins.

We fit the profiles of the $\Delta$[Fe/H]/$\Delta|Z|$--$R$ for the thin/thick disks (see Fig.\,7), which yields  $\Delta$[Fe/H]/$\Delta|Z|$ $\propto$ $R^{-2.307}$ and $\Delta$[Fe/H]/$\Delta|Z|$ = 0.001 $R$ $-$ 0.078 for the thin and thick disks, respectively (to distinguish it from the original data, hereafter, we name the profiles of $\Delta$[Fe/H]/$\Delta|Z|$ as Fit\_$\Delta$[Fe/H]/$\Delta|Z|$).
After subtracting the fitted profile of the original data, we present the correction vertical metallicity gradient (Cor\_$\Delta$[Fe/H]/$\Delta|Z|$ = $\Delta$[Fe/H]/$\Delta|Z|$ $-$ Fit\_$\Delta$[Fe/H]/$\Delta|Z|$) as a function of $R$ in Fig.\,8.
The result indicates the Cor\_$\Delta$[Fe/H]/$\Delta|Z|$ exhibits some oscillation with $R$ for the thin and thick disks.
The resonances with the Galactic bar are the most possible reason for these oscillations \citep[e.g.,][]{Kalnajs1991, Dehnen2000, Antoja2014, Hunt2018, Sun2023}.

A north-south asymmetry in [Fe/H] is obvious for the thin disk at $R$ $>$ 9.0\,kpc (see Fig.\,5), which is in rough agreement with the warp in structure \citep[e.g.,][]{Momany2006, Poggio2018, Mackereth2019, Li2020}.
In further work, we will attempt to establish a correlation between the north-south asymmetry in [Fe/H] and the disk warp, and the related study is underway and shall be presented in a separate paper (Sun et al. {\color{blue}{2024}}, in preparation).
The north-south asymmetry in [Fe/H] signal is weak for the thick disk (see Fig.\,5 and upper right panel of Fig.\,7), and such asymmetry signal is further weakened but did not disappear when we limit the [$\alpha$/Fe] of thick disk stars larger than 0.15\,dex.
This implies that the thick disk may also have a weak warp signal if the north-south chemical asymmetry is caused by the Galactic warp \citep[e.g.,][]{Li2020, Sun2024}.
The weak signal for the thick disk is in good agreement with the old/high-$\alpha$ populations in recent studies \citep[see e.g.,][]{Poggio2018, Sun2024}.
This indicates that the metallicity asymmetry (and the warp) may be a long-lived feature.

\begin{figure*}[t]
\centering
\subfigure{
\includegraphics[width=8.4cm]{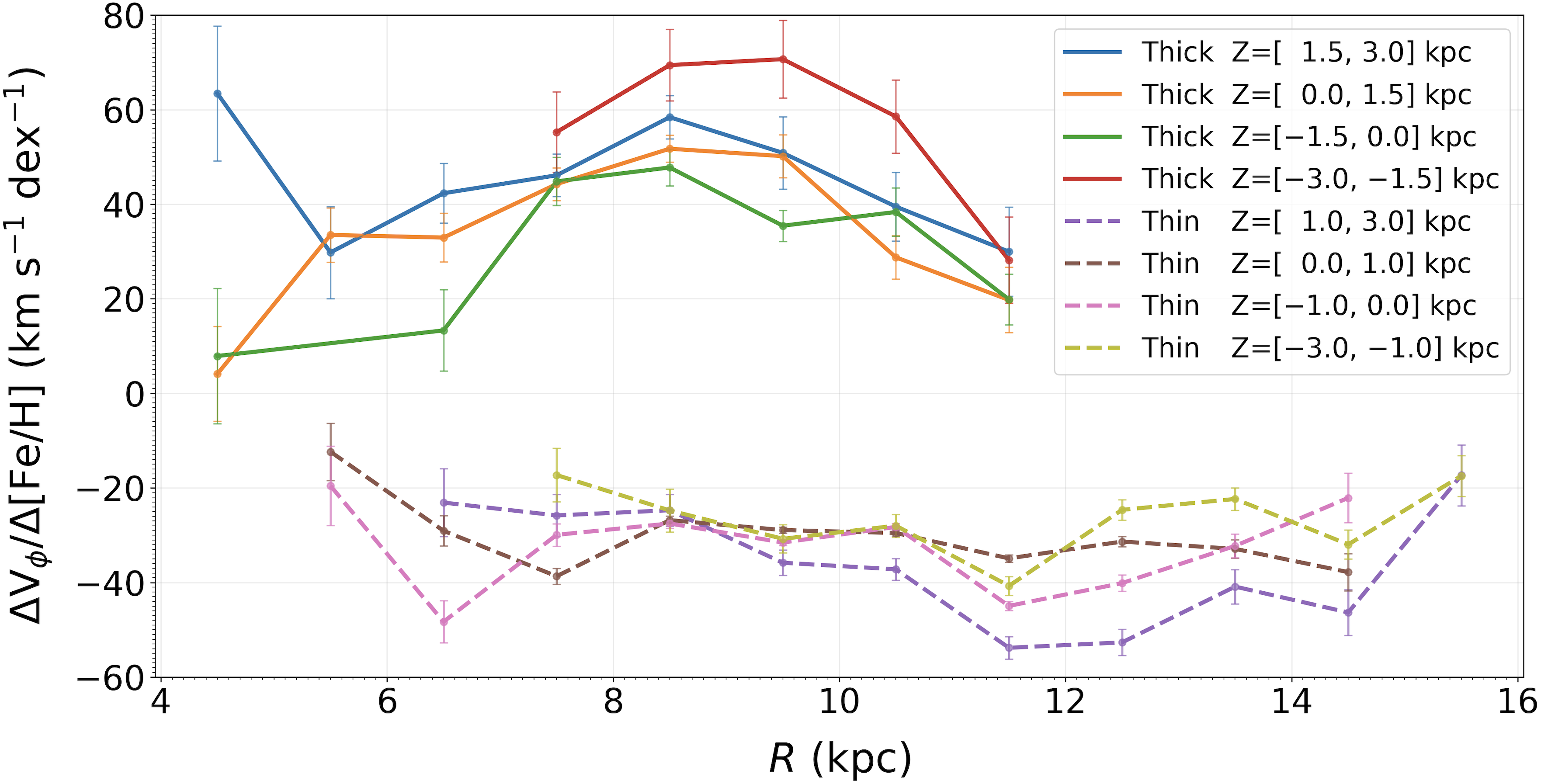}
}
\hspace{0.5cm}
\subfigure{
\includegraphics[width=8.4cm]{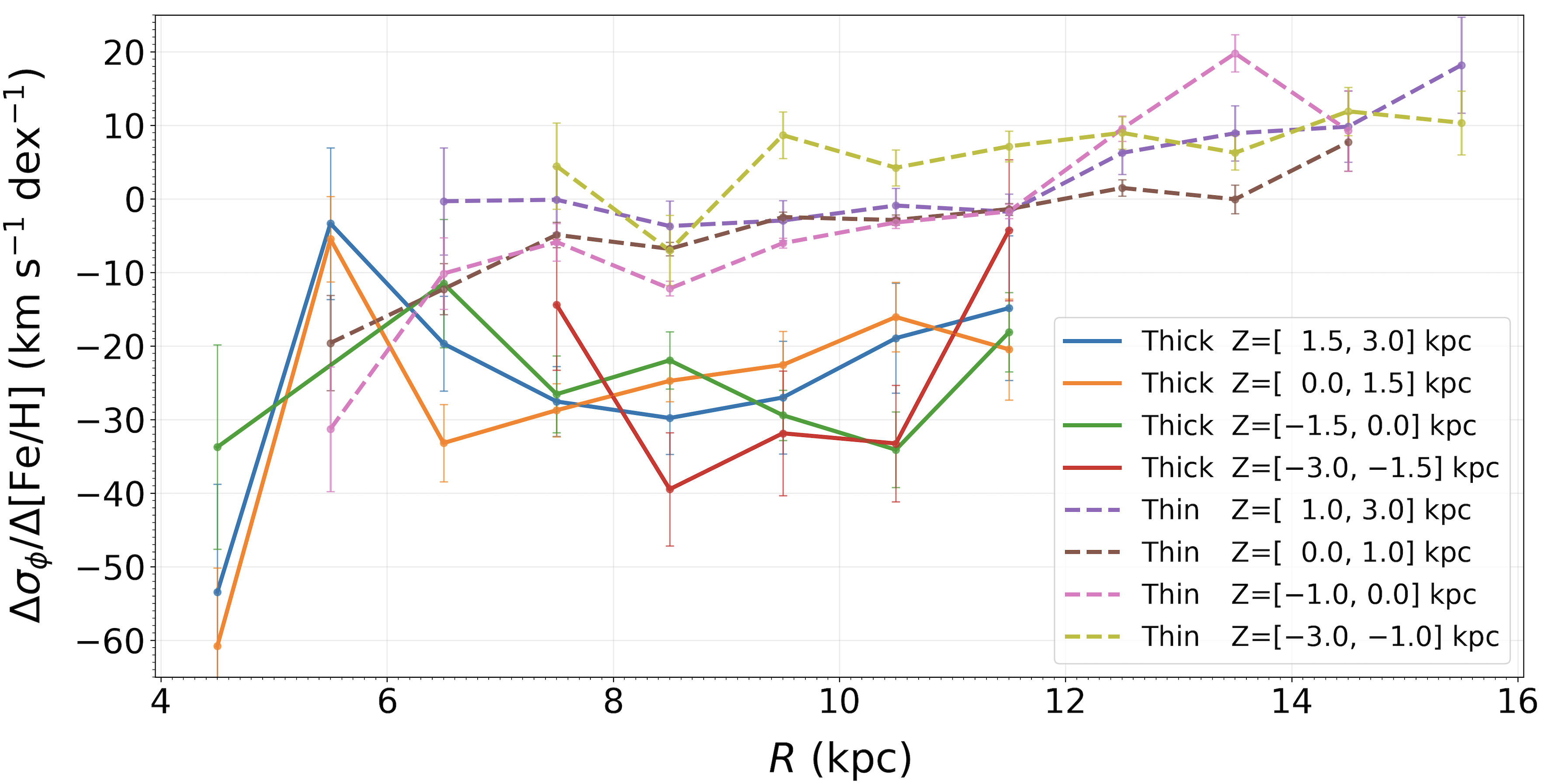}
}

\caption{Metallicity gradients of the azimuthal velocity (left panel) and azimuthal velocity dispersion (right panel) as a function of $R$ and $Z$.}
\end{figure*}
%%\label{fig9}

\begin{figure*}[t]
\centering
\subfigure{
\includegraphics[width=8.4cm]{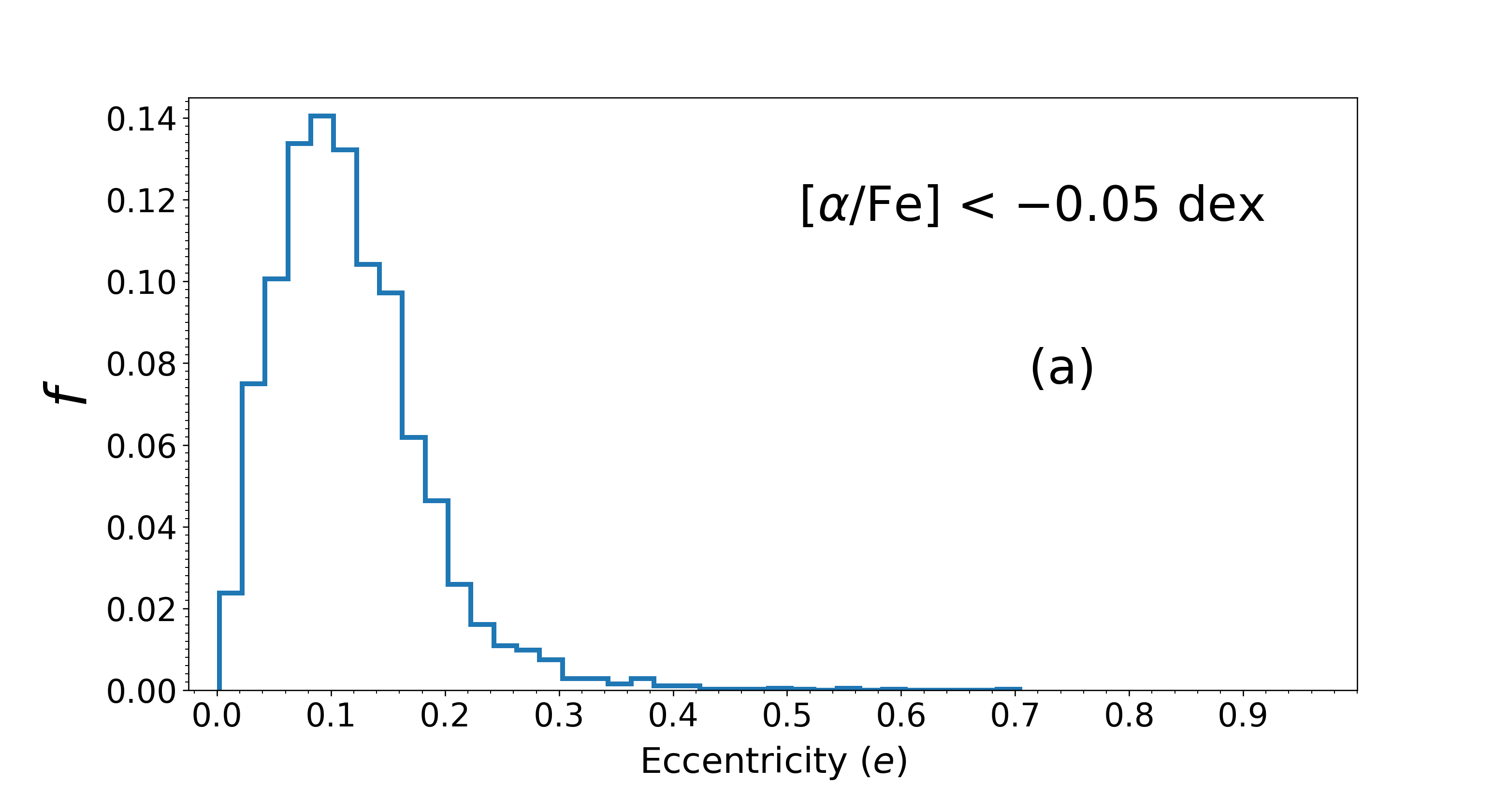}
}
\subfigure{
\includegraphics[width=8.4cm]{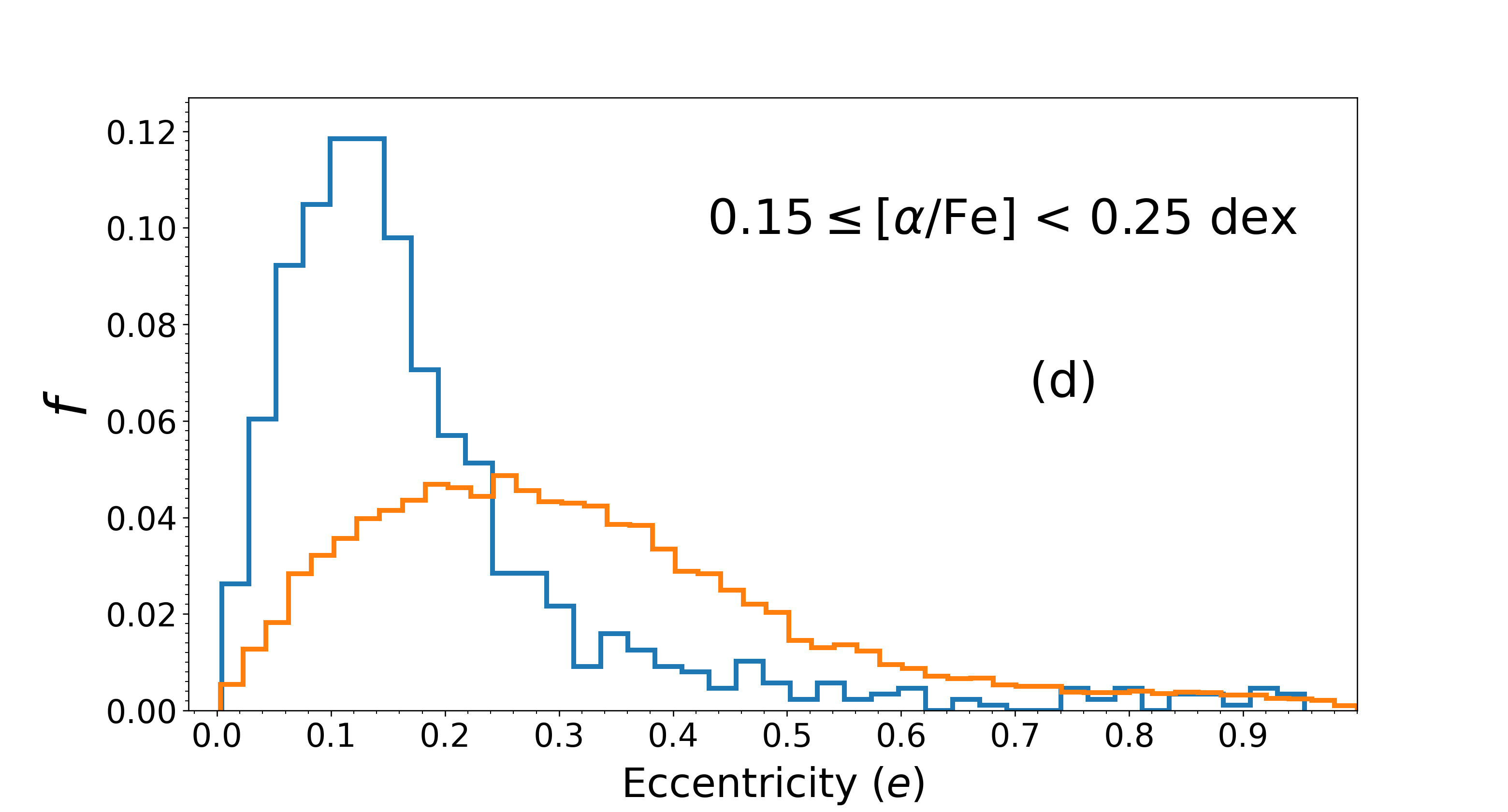}
}

\subfigure{
\includegraphics[width=8.4cm]{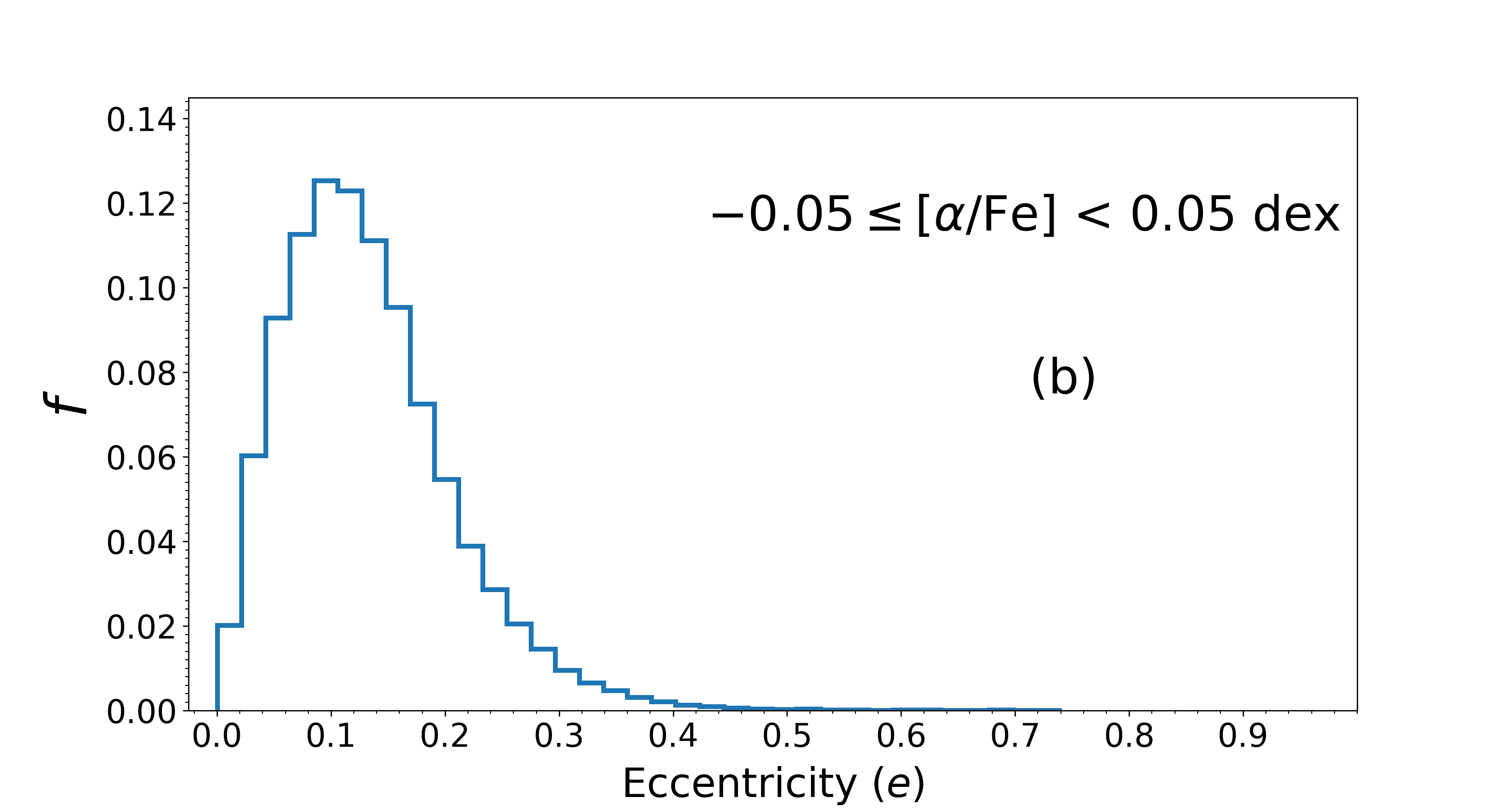}
}
\subfigure{
\includegraphics[width=8.4cm]{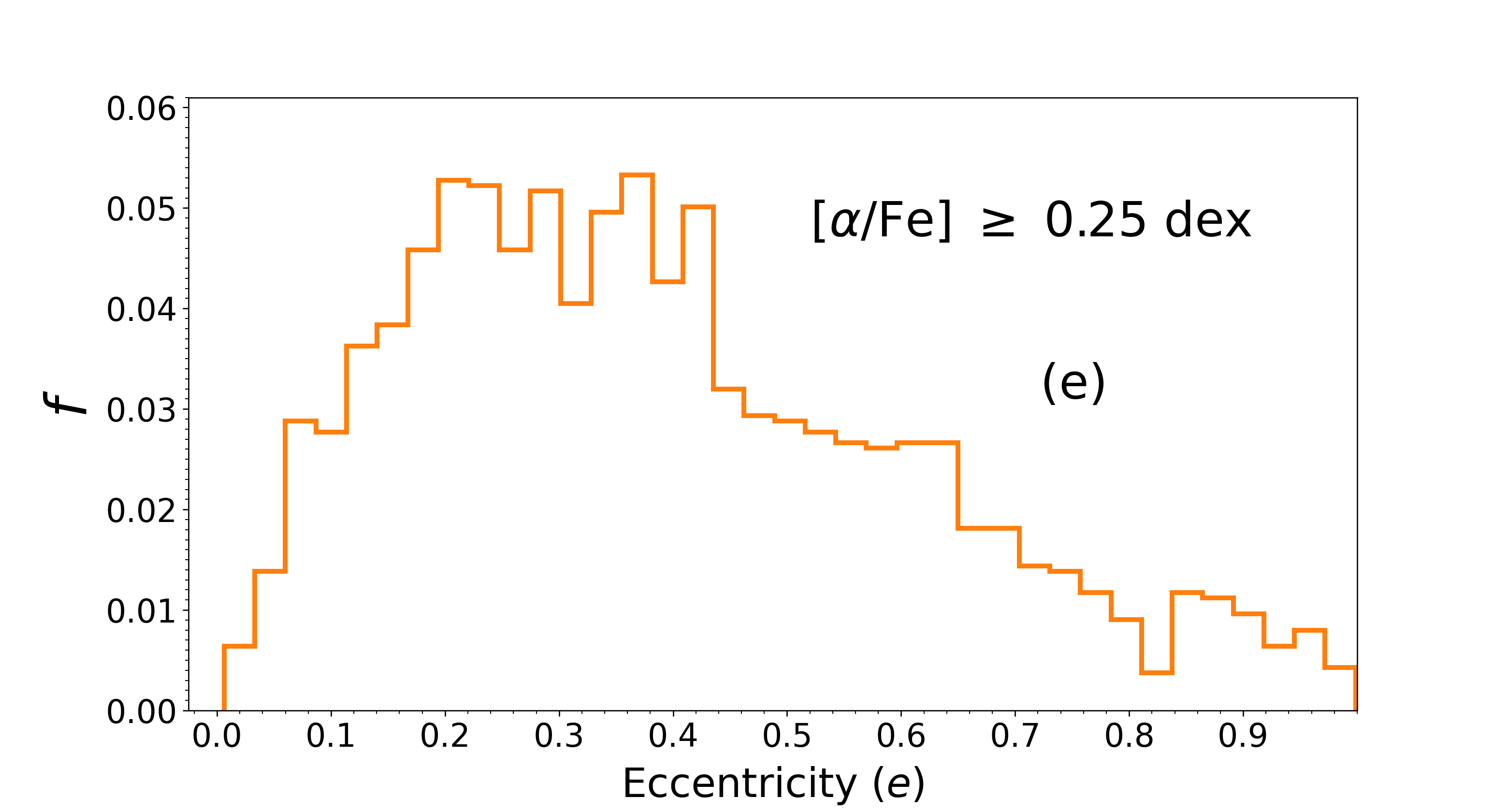}
}

\subfigure{
\includegraphics[width=8.4cm]{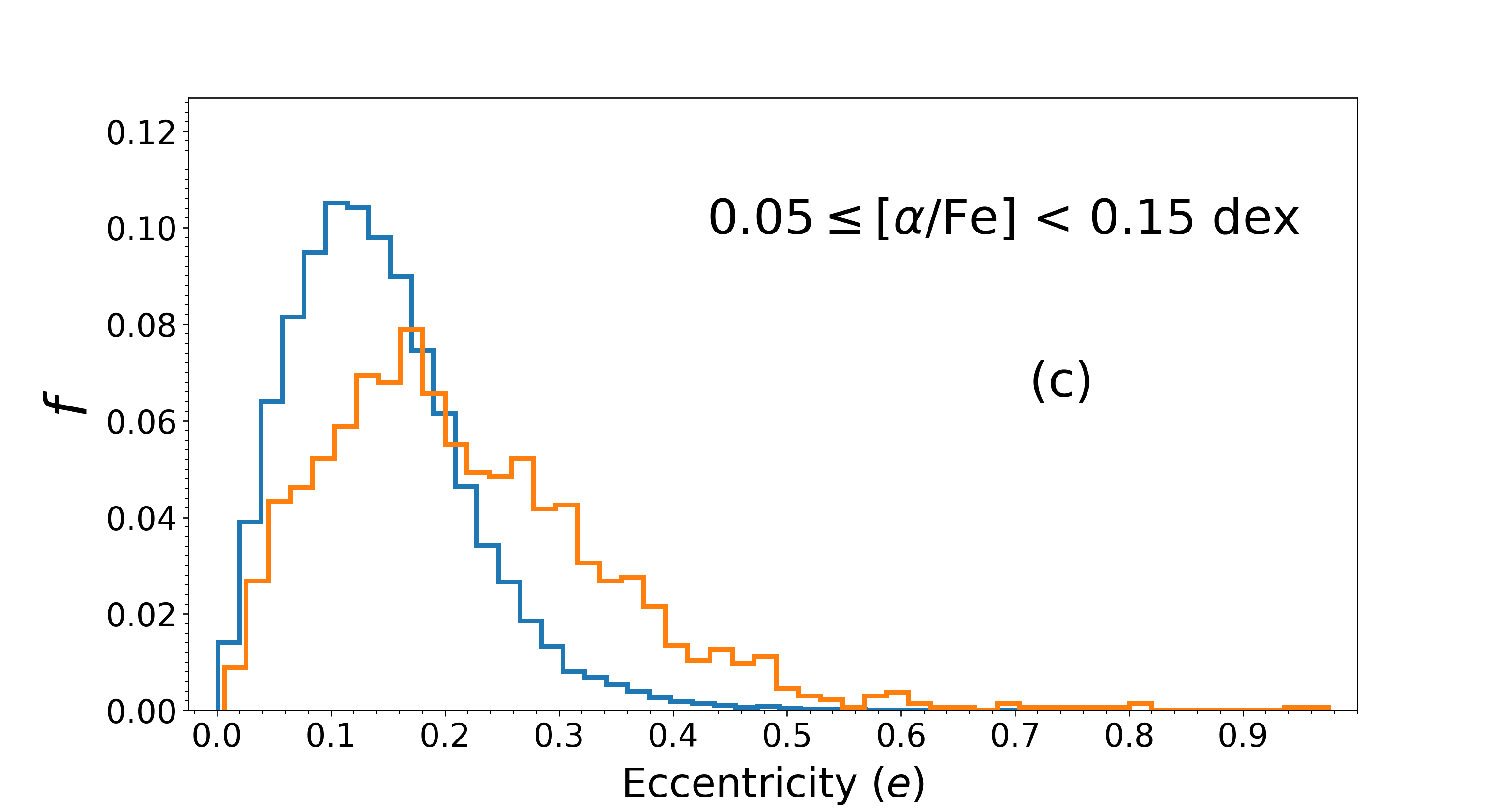}
}
\subfigure{
\includegraphics[width=8.4cm]{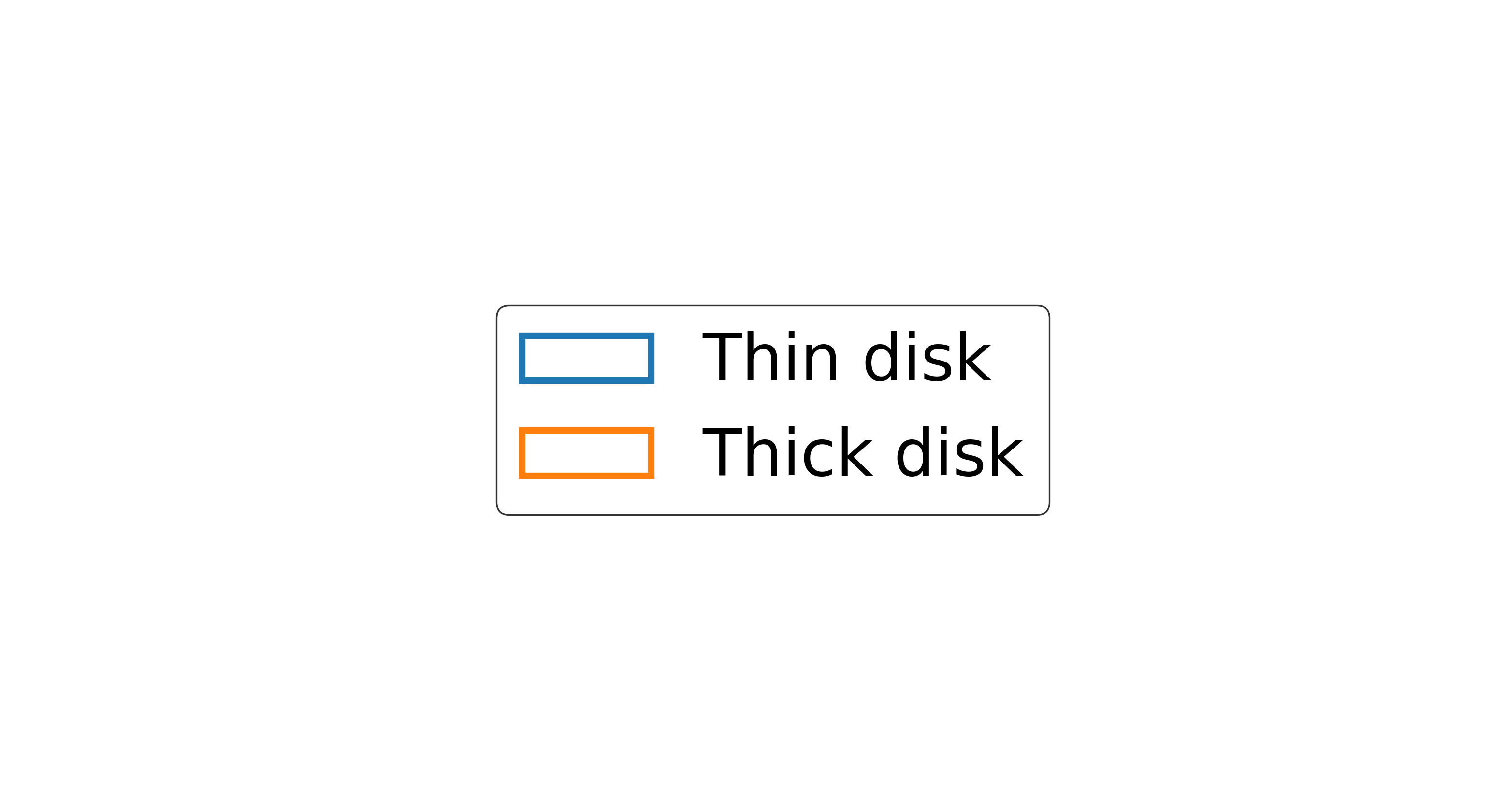}
}
\caption{Histograms of eccentricities for the thin-disk (black line) and thick-disk (red line) stars in different [$\alpha$/Fe] bins, as indicated in the legends of each panel. The [$\alpha$/Fe] range increases from panels (a) to (e).}
\end{figure*}
%%\label{fig10}

\section{The $\Delta$V$_{\phi}$/$\Delta$[Fe/H] and $\Delta \sigma_{\phi}$/$\Delta$[Fe/H] distributions of the thin/thick disks}

The metallicity gradients of azimuthal velocity and its dispersion provide a more intuitive illustration of the stellar radial migrations \citep[e.g.,][]{Lee2011, Han2020}.
Here, we determined the $\Delta$V$_{\phi}$/$\Delta$[Fe/H] and $\Delta \sigma_{\phi}$/$\Delta$[Fe/H] as a function of $R$ and $Z$ for the thin and thick disks (see Fig.\,9).

For the thin disk, $\Delta$V$_{\phi}$/$\Delta$[Fe/H] shows a global negative value for all $R$ bins.
In the solar radius (8.0 $<$ R $\leq$ 9.0\,kpc), $\Delta$V$_{\phi}$/$\Delta$[Fe/H] is around $-$27.48 $\pm$ 1.00\,km\,s$^{-1}$\,dex$^{-1}$ at $-$1.0 $\leq$ Z $<$ 0.0\,kpc, and is around $-$26.81 $\pm$ 0.91\,km\,s$^{-1}$\,dex$^{-1}$ at 0.0 $\leq$ Z $<$ 1.0\,kpc.
Our result $\Delta$V$_{\phi}$/$\Delta$[Fe/H] = $-$27.48 $\pm$ 1.00\,km\,s$^{-1}$\,dex$^{-1}$ is in good agreement with the result of Han et al. ({\color{blue}{2020}}), who reported that $\Delta$V$_{\phi}$/$\Delta$[Fe/H] is $-$28.2\,km\,s$^{-1}$\,dex$^{-1}$ for the G- and K-dwarfs in the solar neighborhood.
The $\Delta \sigma_{\phi}$/$\Delta$[Fe/H] shows a slight increasing trend with $R$, rising from $\Delta \sigma_{\phi}$/$\Delta$[Fe/H] $\sim$ $-$10.0\,km\,s$^{-1}$\,dex$^{-1}$\,kpc$^{-1}$ at $R$ $\sim$\,7.0\,kpc to $\Delta \sigma_{\phi}$/$\Delta$[Fe/H] $\sim$ 20.0\,km\,s$^{-1}$\,dex$^{-1}$\,kpc$^{-1}$ at $R$ larger than $\sim$\,15.0\,kpc.
For the thick disk, $\Delta$V$_{\phi}$/$\Delta$[Fe/H] and $\Delta \sigma_{\phi}$/$\Delta$[Fe/H] are respectively positive and negative.
The $\Delta \sigma_{\phi}$/$\Delta$[Fe/H] for the thick disk is globally stronger than that for the thin disk.

The $\Delta$V$_{\phi}$/$\Delta$[Fe/H] is respectively negative and positive for thin and thick disks is interesting.
To explain this behavior, it is necessary to clarify the meaning of churning and blurring in the `radial migration' terms in the following analysis.
In churning, the angular momentum of stellar orbit changes due to the interaction of the Galactic non-asymmetric structure (i.e., the Galactic bar or spiral arms), and the change of the stellar angular momentum leads to the change of the stellar guiding radius, and so as leads to stellar radial migration \citep[e.g.,][]{Sellwood2002, Schonrich2017, Hayden2020}.
In blurring, stellar kinematics become heated (by scattering of the GMCs, and resonance between the Galactic bar and spiral arms) with time, resulting in stellar radial velocity dispersion rising with time, and the larger radial velocity dispersion of stars leads to large epicyclic motions over their orbits with the angular momentum of each star conserved \citep[e.g.,][]{Minchev2010, Schonrich2017, Hayden2020}.

The negative $\Delta$V$_{\phi}$/$\Delta$[Fe/H] of the thin disk is widely reported by previous studies in the solar neighborhood \citep[e.g.,][]{Spagna2010, Lee2011, Recio-Blanco2014, Kordopatis2017, Hayden2020, Han2020}.
Due to the negative radial metallicity gradient of the thin disk stars (see Fig.\,5-7), stars with richer metallicity generally have smaller Galactic guiding radii and move outward with a blurring radial mixing mechanism that leads these stars to be observed with lower azimuthal velocities than local stars,
while stars formed in the outer Galactic disk generally have lower metallicity and higher azimuthal velocities than those of local stars, thus resulting in negative $\Delta$V$_{\phi}$/$\Delta$[Fe/H].

For the thick disk stars, the inside-out and upside-down formation of the thick disk \cite[e.g.,][]{Kawata2017, Schonrich2017}, along with the stellar radial migration \citep[e.g.,][]{Zhang2020, Han2020}, can easily explain the observed results.
As the radial velocity dispersion is quite large for thick disk stars \citep[e.g.,][]{Lee2011, Hayden2020, Sun2024}, these stars experience obvious radial epicyclic motion which results in lower $V_{\phi}$ as they move outwards.
As the radial velocity dispersion is smaller for the more metal-rich thick disk stars than that for the metal-poor stars \citep[e.g.,][]{Hayden2020, Sun2024}, this causes smaller radial epicyclic motion and larger $V_{\phi}$ for the metal-rich stars than those for the metal-poor stars.
Combined with stellar positive radial metallicity gradient (see Fig.\,6--7), we can well explain the relations of the metallicity with $V_{\phi}$ and $\sigma_{\phi}$.

The V$_{\phi}$/$\Delta$[Fe/H] of the thin disk shows a global slight increasing trend with $R$ at $R$ $>$ 11.5\,kpc (see left panel of Fig.\,9).
We consider some possible explanations that can explain this behavior:
(i) Stars in this region are generally dominated by migrators that formed in the inner disk, metal-rich stars generally have experienced larger radial epicyclic motion than those of metal-poor stars, and so as lead to a strong gradient;
(ii) churning as the likely mechanism of radial mixing for a significant number of stars in this region;
(iii) the flared nature \citep[e.g.,][]{Minchev2015, Bovy2016, Sun2020, Vickers2021} in the outer Galactic disk.
The increasing trend of $\Delta \sigma_{\phi}$/$\Delta$[Fe/H]--$R$ for the thin disk (see right panel of Fig.\,9), may be linked to the warped disk.

\section{Eccentricity distributions of the thin/thick disks}

Different dynamic heating processes can be clearly detected in the properties of chemo-dynamics \citep[e.g.,][]{Quinn1993, Abadi2003, Brook2004, Sales2009, Schonrich2009, Loebman2011, Minchev2012, Schonrich2017}.
Here, we present the distributions of the eccentricities of different [$\alpha$/Fe] populations in Fig.\,10.
For the thin disk, all the various [$\alpha$/Fe] sub-populations peak around 0.1 in eccentricities, with similar distribution shapes.
It is somewhat surprising that the thin disk with 0.15 $\leq$ [$\alpha$/Fe] $<$ 0.25 in panel (d) seems to have some stars with high eccentricities (typically larger than $\sim$0.4).
The ``young” [$\alpha$/Fe]-enhanced stars \citep[e.g.,][]{Chiappini2015, Martig2015, Jofre2016, Yong2016, Matsuno2018, Hekker2019, Sun2020} can explain some of these stars, and the contamination of the thick disk stars also would contribute a substantial amount.

For the thick disk, the eccentricity distributions are distinguished for different [$\alpha$/Fe] sub-populations in that the 0.05 $\leq$ [$\alpha$/Fe] $<$ 0.15, 0.15 $\leq$ [$\alpha$/Fe] $<$ 0.25 and [$\alpha$/Fe] $\geq$ 0.25 populations are respectively peaked around, 0.18, 0.25 and 0.3.
In panel (e), the distributions seem to show weak double peaks, with the second peak at around 0.9.
In comparison with the simulations of Sales et al. ({\color{blue}{2009}}) who predicted the distributions of the eccentricities by heating, accretion, migration and merger,
our results indicate that the thick disk stars in panels (c), (d) and (e), are respectively, mainly heated by migration, gas-rich mergers and accretion.
In addition, the heating by migration is indispensable for both thin and thick disks.

\section{Discussion on the formation and evolution of the Galactic disk}

These distinguished properties of the thin and thick disks have been widely measured and characterized by previous studies \citep[e.g.,][]{Fuhrmann1998, Bensby2005, Lee2011, Yan2019, Han2020, Sun2024}.
These results indicate the thin and thick disks may have experienced different formation mechanisms and/or evolution histories \citep[e.g.,][]{Jenkins1992, Brook2004, Mackereth2019, Han2020}.
Briefly summarizing, the thin disk is widely considered to originate from dynamic heating by both GMCs and spiral arms \citep[e.g.,][]{Spitzer1953, Jenkins1992, Mackereth2019}, and the disk warp, flare, bending, and some perturbation events also contributed an amount for the heating of the thin disk stars \citep[e.g.,][]{Khoperskov2017, Mackereth2019, Sun2024}.
For thick disk, the ``inside-out" and ``upside-down" stars formation, along with radial migration, can readily understand the observed properties  \citep[e.g.,][]{Schonrich2017, Vickers2021, Han2020}.
Several studies also revealed that the thick disk stars also require a significant contribution from some violent heating mechanisms, such as merger \citep[e.g.,][]{Quinn1993, Villalobos2008} and accretion \citep[e.g.,][]{Abadi2003}, or born in the chaotic mergers of gas-rich systems and/or turbulent interstellar medium \citep[e.g.,][]{Brook2004, Wisnioski2015}.
The ``two-infall" \citep[e.g.,][]{Chiappini1997, Spitoni2019, Lian2020} and a clumpy stars formation \citep[e.g.,][]{Bournaud2007, Clarke2019, Amarante2020} models can also explain the thick disk stars.

Since the existing simulations/models are affected by many factors, i.e., some assumptions and unavoidable numerical bias, the simulations/models for the formation and evolution of the disk may be an incomplete physical reality, and hence, the comparison of our results with simulations/models cannot be well quantized.
Therefore, we only compare qualitatively our results with expectations from the results of the simulations/models that are related to radial migration, merger, accretion, gas-rich systems and other disk heating simulations/models.

\begin{figure*}[t]
\centering
\subfigure{
\includegraphics[width=8.6cm]{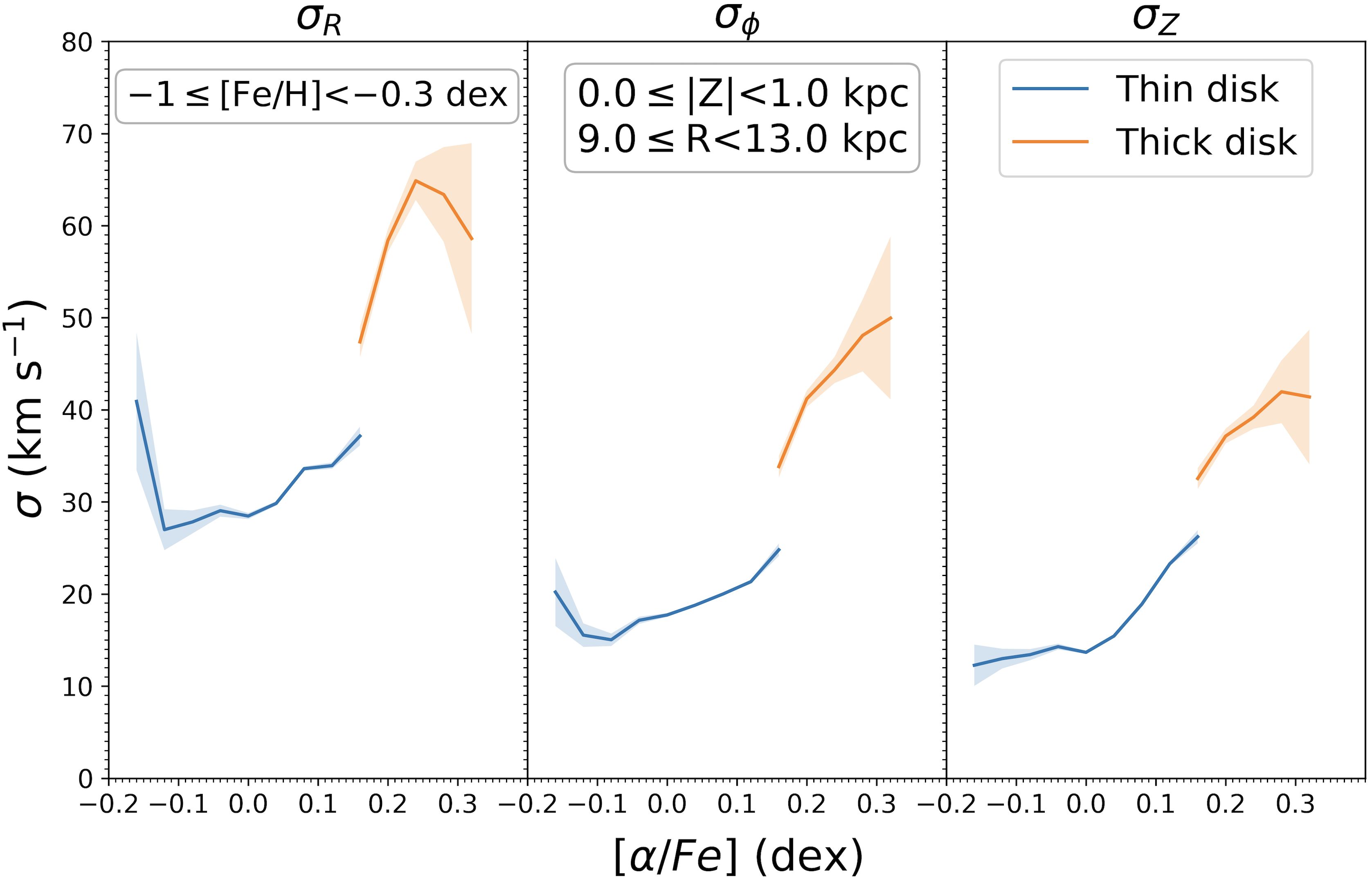}
}
\hspace{0.4cm}
\subfigure{
\includegraphics[width=8.6cm]{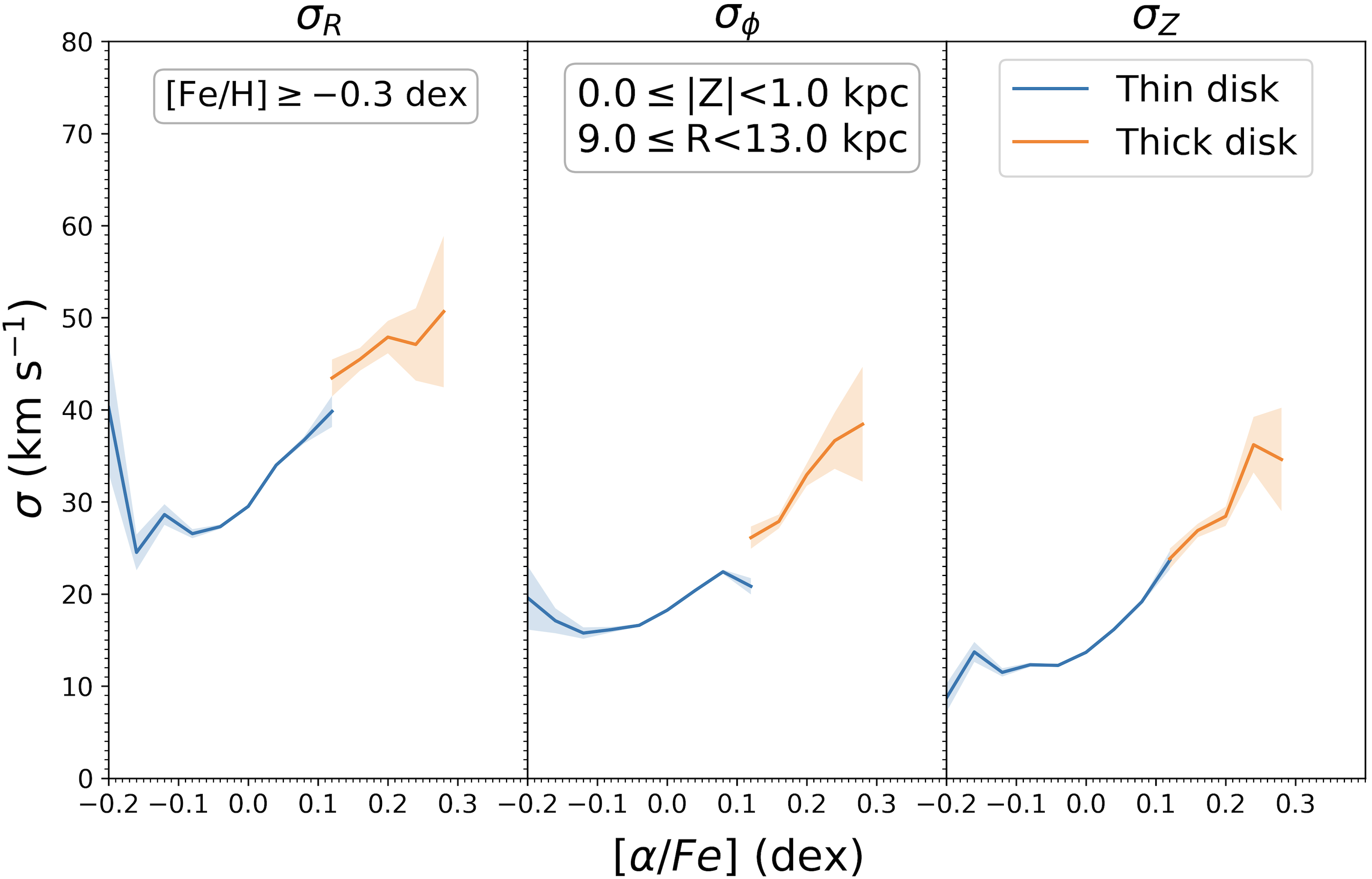}
}

\caption{The [$\alpha$/Fe]--velocity dispersion for the metal-poor (left panel) and metal-rich (right panel) populations.}
\end{figure*}
%%\label{fig11}

For the thin disk, the radial migration simulations/models \citep[e.g.,][]{Sellwood2002, Schonrich2009, Minchev2010, Schonrich2017} predict that gas density in the inner disk is higher (therefore the [Fe/H] is richer) than the outer disk for the thin disk.
Thus, stars born in the inner disk are richer in [Fe/H] than in the outer disk.
This means the thin disk has an initial negative metallicity gradient with $R$ in the earlier times.
These stars could be changed by radial migration, which results in a flat radial metallicity gradient.
Loebman et al. ({\color{blue}{2011}}) reported a weak radial metallicity with $\Delta$[Fe/H]/$\Delta$R = $-$0.02\,dex\,kpc$^{-1}$ for the thin disk in the solar neighborhood ($R$ = 7--11\,kpc and $Z$ = 0.3--2.0\,kpc), by using N-body simulations of radial migration.
This result is in good agreement with ours for the thin disk in the north with $Z$ = 1.5 $\sim$ 2.0\,kpc and in the south with $Z$ = $-$2.5 $\sim$ $-$2.0\,kpc (see Fig.\,7).
Furthermore, Loebman et al. ({\color{blue}{2011}}) also reported the thin disk in the region of $R$ = 7--9\,kpc and $|Z|$ = 0.1--1\,kpc has $\Delta$V$_{\phi}$/$\Delta$[Fe/H] = $-$19.03\,km\,s$^{-1}$\,dex$^{-1}$.
This result is weaker than our gradient with $\Delta$V/$\Delta$[Fe/H] = $-$26.81 $\pm$ 0.91\,km\,s$^{-1}$\,dex$^{-1}$ (see Fig.\,9), while our result is well consistent with the result from Han et al. ({\color{blue}{2020}}) with $\Delta$V/$\Delta$[Fe/H] = $-$27.48 $\pm$ 1.00\,km\,s$^{-1}$\,dex$^{-1}$.
Thus, we consider that the $\Delta$[Fe/H]/$\Delta$R of the thin disk can be well explained by the global radial migration simulations/models, while our result of $\Delta$V/$\Delta$[Fe/H] of the thin disk indicates that the radial migration may be only one of the multiple contributions.
The difference of the relation of $\Delta$[Fe/H]/$\Delta$R vs. $|Z|$ (see upper left panel of Fig.\,7) shows that the $\Delta$[Fe/H]/$\Delta$R of the thin disk is obvious stronger for the south disk than that for the north disk, indicating that the disk may have experienced some turbulent events from south to north \citep[e.g.,][]{Gomez2012a, Gomez2012b, Laporte2018, Carrillo2019, Sun2024}, and/or may be linked to the disk warp \citep[e.g.,][]{Momany2006, Gaia Collaboration2018}.
The oscillations of the Cor\_$\Delta$[Fe/H]/$\Delta |Z|$ with $R$ for the thin and thick disks may indicate the resonance with the Galactic Bar.
The trend of the $\Delta$[Fe/H]/$\Delta$R with $|Z|$ (and/or the trend of $\Delta$[Fe/H]/$\Delta |Z|$ with $R$) may be linked to a flared disk \citep[e.g.,][]{Minchev2015, Bovy2016, Sun2024}.

For the thick disk, Loebman et al. ({\color{blue}{2011}}) suggested $\Delta$[Fe/H]/$\Delta$R = 0.00 dex\,kpc$^{-1}$ and $\Delta$[Fe/H]/$\Delta|Z|$ = $-$0.03\,dex\,kpc$^{-1}$ by the N-body simulation of radial migration.
Their radial metallicity gradient is well consistent with our results (see upper right panel of Fig.\,7), while their vertical metallicity gradient is slightly smaller than our result of $\Delta$[Fe/H]/$\Delta|Z|$ = $-$0.06 $\sim$ $-$0.08\,dex\,kpc$^{-1}$ (see bottom panel of Fig.\,7).
The suggested $\Delta$V/$\Delta$[Fe/H] as  +8.0\,km\,s$^{-1}$\,dex$^{-1}$ by the simulations/models of radial migration \citep{Loebman2011} is also obviously smaller than our result (see Fig.\,9).
From the simulations/models of Sales et al. ({\color{blue}{2009}}), the fraction of high eccentricity (e $>$ 0.5) thick disk stars formed by radial migration heating mechanism is apparently less than our results as exhibited in Fig.\,10.
These results indicate that the radial migration contributes just some amount to the formation and evolution of the thick disk, but does not dominate.
Thus, some additional star formation and heating mechanisms of the thick disk may be more in line with our measurements.
The global positive $\Delta$V$_{\phi}$/$\Delta$[Fe/H] and $\Delta$[Fe/H]/$\Delta$R of the thick disk (Fig.\,8--9) may be some indication of the ``inside-out" and ``upside-down" star formation scenario \citep[e.g.,][]{Kawata2017, Schonrich2017}.
The eccentricity distributions indicate that thick disk stars have experienced some violent heating processes, such as migration, merger and accretion (see Fig.\,10).
The trend of the $\Delta$V$_{\phi}$/$\Delta$[Fe/H] (and/or $\Delta \sigma_{\phi}$/$\Delta$[Fe/H]) with $R$ may indicate a warped thick disk \citep[e.g.,][]{Li2020, Sun2024}.

The [$\alpha$/Fe]--[Fe/H] distribution in our result indicates that the thin and thick disks are non-continuous.
This is in good agreement with previous studies \citep[e.g.,][]{Fuhrmann1998, Lee2011, Han2020, Sun2020}, and to some extent opposes the continuous disk model \citep[e.g.,][]{Bensby2007, Bovy2012a, Kawata2016, Hayden2017}.
Our results with non-continuous [$\alpha$/Fe]--[Fe/H] distribution of the thin and thick disk is well consistent with the predictions of the ``two-infall" \citep[e.g.,][]{Chiappini1997, Spitoni2019, Lian2020} and the clumpy stars formation \citep[e.g.,][]{Bournaud2007, Clarke2019, Amarante2020} models.
Our measurement radial and vertical profiles of the thick disk (see left panels of Fig\,6) are also in line with the results of the clumpy stars formation model in Clarke et al. ({\color{blue}{2019}}).

The stellar [$\alpha$/Fe]--velocity dispersion ($\sigma$) relations are well characterized in the solar neighborhood \citep[e.g.,][]{Minchev2013, Minchev2014, Hayden2020}, which provide rich information for the perturbation and heating histories of the Galactic disk \citep[e.g.,][]{Minchev2014, Hayden2020, Sun2024}
The global increasing trends of the [$\alpha$/Fe]--$\sigma$ are clearly detected for various populations (see Fig.\,11), which are likely because of the heating from Giant Molecular Clouds (GMCs) \citep[e.g.,][]{Spitzer1951, Spitzer1953, Barbanis1967, Jenkins1992}.
The global trend of $\sigma_{R}$ $>$ $\sigma_{\phi}$ $>$ $\sigma_{Z}$ at the same [$\alpha$/Fe] (see Fig.\,11), may point to the heating from transient spiral arms \citep[e.g.,][]{Jenkins1992, Sun2024}.

The trend of the [$\alpha$/Fe]--$\sigma$ relation for the thick disk deviates significantly from the growth trend of the [$\alpha$/Fe]--$\sigma$ relation for the thin disk, showing an obvious gap for the thin and thick disks (see Fig.\,11).
This result indicates that the thick disk stars may have experienced some violent heating processes, such as the minor mergers, accretions and infall of misaligned gas \citep[e.g.,][]{Quinn1993, Abadi2003, Roskar2010, Grand2016, Kazantzidis2008, Aumer2013, Belokurov2018, Deason2018, Helmi2018, Kruijssen2019, Sun2024}.
The thick disk shows a flat/weak trend in $\sigma$ as [$\alpha$/Fe] increases at [$\alpha$/Fe] $>$ 0.2\,dex, and different [$\alpha$/Fe] shows a similar velocity dispersion, which implies that these thick disk stars are possibly formed in the chaotic merger of gas-rich systems and turbulent interstellar medium (ISM) \citep[e.g.,][]{Brook2004, Brook2007, Brook2012, Wisnioski2015, Mackereth2019, Sun2024}

As a reminder of the influence of the data uncertainties on our results, although we have tested the reliability of the results in this paper, the impact on the understanding of the assembly history of the Milky Way still can not be completely ruled out.
A later complete physical reality simulations/models and higher-precision observations may help us to make a complete understanding of the formation and evolution of the Galactic disk.

\section{Conclusions}

Using over 170,000 red clump stars selected from the LAMOST and APOGEE spectroscopic surveys, we make a detailed measurement of the metallicity distributions, kinematics and dynamics of the thin and thick disks mainly between 5.0 $\leq$ $R$ $\leq$ 15.0\,kpc and $|Z|$ $\leq$ 3.0\,kpc with high accuracy.
We find that:
\\
\\
$\bullet$ Thin and thick disks display significant differences in the metallicity, $\Delta$V$_{\phi}$/$\Delta$[Fe/H], $\Delta \sigma_{\phi}$/$\Delta$[Fe/H] and eccentricity distributions, as well as the [$\alpha$/Fe]--velocity dispersion relations.
The spatial variations in $\Delta$[Fe/H]/$\Delta$R and $\Delta$[Fe/H]/$\Delta |Z|$ of the thin disk are globally stronger than those of the thick disk.
The $\Delta$V$_{\phi}$/$\Delta$[Fe/H] is respectively negative and positive for the thin and thick disks, and the $\Delta \sigma_{\phi}$/$\Delta$[Fe/H] of the thin disk is globally weaker than the thick disk.
The north-south asymmetry in [Fe/H] is significant for the thin disk, whereas the thick disk shows much weaker asymmetry.
The [$\alpha$/Fe]--velocity dispersion relation of the thin disk shows an increasing trend with [$\alpha$/Fe] at its high [$\alpha$/Fe] ($>$ 0.05\,dex), whereas the thick disk shows a flat trend at its high [$\alpha$/Fe] ($>$ 0.2\,dex).
\\
\\
$\bullet$ The metallicity distributions indicate that the thin disk is generally steeper in $\Delta$[Fe/H]/$\Delta$R than the thick disk that the $\Delta$[Fe/H]/$\Delta$R shows an increasing trend with $|Z|$, rising from $\Delta$[Fe/H]/$\Delta$R $\sim$ $-$0.06\,dex\,kpc$^{-1}$ at $|Z|$ $<$ 0.25\,kpc to $\Delta$[Fe/H]/$\Delta$R $\sim$ $-$0.02\,dex\,kpc$^{-1}$ at $|Z|$ $>$ 2.75\,kpc.
The thick disk shows a global weak positive $\Delta$[Fe/H]/$\Delta$R, generally weaker than 0.01\,dex\,kpc$^{-1}$, and this gradient shows no obvious change with $|Z|$.
The $\Delta$[Fe/H]/$\Delta|Z|$ of the thin disk follows a single power law that increases steadily from around $-$0.36\,dex\,kpc$^{-1}$ at $R$ $\sim$ 5.5\,kpc to around $-$0.05\,dex\,kpc$^{-1}$ at R $>$ 11.5\,kpc.
The thick disk exhibits an almost constant gradient (nearly $-$0.06 $\sim$ $-$0.08\,dex\,kpc$^{-1}$) for all R bins.
The north-south asymmetry in [Fe/H] may be linked to the disk warp and/or the disk perturbation events, and the trend of the $\Delta$[Fe/H]/$\Delta$R with $|Z|$ (and/or the trend of $\Delta$[Fe/H]/$\Delta |Z|$ with R) may indicate that the disk warp \citep[e.g.,][]{Momany2006, Poggio2018} and flare \citep[e.g.,][]{Minchev2015, Bovy2016}, as well as the oscillations of the Cor\_$\Delta$[Fe/H]/$\Delta|Z|$ with $R$ for the two disks may be linked to the resonance with the Galactic Bar \citep[e.g.,][]{Kalnajs1991, Dehnen2000}.
\\
\\
$\bullet$ The $\Delta$V$_{\phi}$/$\Delta$[Fe/H] is respectively negative and positive for the thin and thick disks, which may indicate an ``inside-out" and ``upside-down" \citep[e.g.,][]{Kawata2017, Schonrich2017} star formation mechanism of the thick disk.
The $\Delta$V$_{\phi}$/$\Delta$[Fe/H] in our measurement is generally stronger than the predicted results of the radial migration simulations/models \citep[e.g.,][]{Sellwood2002, Schonrich2017}, which indicates that the disk stars may have experienced some turbulent events or have experienced some violent heating processes.
\\
\\
$\bullet$ The eccentricity results indicate that the thick stars with 0.05 $\leq$ [$\alpha$/Fe] $<$ 0.15\,dex, 0.15 $\leq$ [$\alpha$/Fe] $<$ 0.25\,dex and [$\alpha$/Fe] $\geq$ 0.25\,dex, may have experienced the heating by migration \citep[e.g.,][]{Sellwood2002, Schonrich2017}, merger \citep[e.g.,][]{Quinn1993, Villalobos2008} and accretion \citep[e.g.,][]{Abadi2003}, respectively.
In addition, both thin and thick disks require a long-term heating by migration.
\\
\\
$\bullet$ The [$\alpha$/Fe]--velocity dispersion relations show a global increasing trend, which indicates the dynamical heating of disk stars by scattering of GMCs and spiral arms \citep[e.g.,][]{Spitzer1951, Barbanis1967, Jenkins1992}.
A gap detected in the relations of [$\alpha$/Fe]--velocity dispersion of thin and thick disks is likely because the thick disk stars may be heated by merger and accretion, or the thick disk stars were born in the chaotic mergers of gas-rich systems and/or turbulent ISM \citep[e.g.,][]{Brook2004, Wisnioski2015}.

\section*{Acknowledgements}

We thank the anonymous referee for very useful suggestions to improve the work.
This work is supported by the NSFC projects 12133002, and National Key R\&D Program of China No. 2019YFA0405503, and CMS-CSST-2021-A09, and National Natural Science Foundation of China grants 12073070, and 12003027.

Guoshoujing Telescope (the Large Sky Area Multi-Object Fiber Spectroscopic Telescope LAMOST) is a National Major Scientific Project built by the Chinese Academy of Sciences. Funding for the project has been provided by the National Development and Reform Commission. LAMOST is operated and managed by the National Astronomical Observatories, Chinese Academy of Sciences. The LAMOST FELLOWSHIP is supported by Special Funding for Advanced Users, budgeted and administrated by Center for Astronomical Mega-Science, Chinese Academy of Sciences (CAMS)

\bibliographystyle{aasjournal}

%\appendix
%\section{Radial metallicity profiles of the thin and thick disks}

\end{document}